\documentclass[onecolumn,showpacs,amsmath,amssymb,eqsecnum]{revtex4}
\usepackage{dcolumn}
\usepackage{bm}
\usepackage[dvips]{graphicx}
\usepackage{wrapfig}
\usepackage{psfrag}
\usepackage{lscape}

\usepackage{amsmath,amssymb,bm}
\usepackage[dvips]{graphicx}

\begin{document}

\Large

\newcommand\x{{\parfillskip0pt\par}}
\newcommand\xe{{\parfillskip0pt\par}\vskip\abovedisplayskip\noindent}

\newcommand\lb{{\linebreak}}
\newcommand\noi{{\noindent}}
\newcommand\pb{{\pagebreak}}

\newcommand\ve[1]{{\mathbf{#1}}}

\newcommand{\Z}{\mathbb{Z}}
\newcommand{\R}{\mathbb{R}}
\def\mK{\mathop{{\mathfrak {K}}}\nolimits}
\def\mR{\mathop{{\mathfrak {R}}}\nolimits}
\def\mv{\mathop{{\mathfrak {v}}}\nolimits}
\def\mS{\mathop{{\mathfrak {S}}}\nolimits}
\def\mU{\mathop{{\mathfrak {U}}}\nolimits}
\def\mI{\mathop{{\mathfrak {I}}}\nolimits}
\def\mA{\mathop{{\mathfrak {A}}}\nolimits}
\def\ma{\mathop{{\mathfrak {a}}}\nolimits}
\def\mH{\mathop{{\mathfrak {H}}}\nolimits}
\def\mN{\mathop{{\mathfrak {N}}}\nolimits}
\def\ml{\mathop{{\mathfrak {l}}}\nolimits}
\def\mf{\mathop{{\mathfrak {f}}}\nolimits}
\def\mo{\mathop{{\mathfrak {o}}}\nolimits}
\def\mP{\mathop{{\mathfrak {P}}}\nolimits}
\def\diag{\mathop{\rm diag}\nolimits}
\newcommand{\ccm}{\cal M}
\newcommand{\cE}{\cal E}
\newcommand{\cV}{\cal V}
\newcommand{\cI}{{\cal I}}
\newcommand{\cR}{{\cal R}}
\newcommand{\cK}{{\cal K}}
\newcommand{\cA}{{\cal A}}
\newcommand{\cU}{{\cal U}}
\newcommand{\cD}{{\cal D}}

\newcommand{\oV}{\overline{V}}
\newcommand{\os}{\overline{s}}
\newcommand{\opsi}{\overline{\psi}}
\newcommand{\ov}{\overline{v}}
\newcommand{\oW}{\overline{W}}
\newcommand{\oPhi}{\overline{\Phi}}
\newcommand{\bx}{\bf{x}}
\newcommand{\by}{\bf{y}}

\def\st{\mathop{\rm st}\nolimits}
\def\tr{\mathop{\rm tr}\nolimits}
\def\sign{\mathop{\rm sign}\nolimits}
\def\d{\mathop{\rm d}\nolimits}
\def\const{\mathop{\rm const}\nolimits}
\def\O{\mathop{\rm O}\nolimits}
\def\Spin{\mathop{\rm Spin}\nolimits}
\def\exp{\mathop{\rm exp}\nolimits}
\def\sh{\mathop{\rm sh}\nolimits}
\def\ch{\mathop{\rm ch}\nolimits}
\def\th{\mathop{\rm th}\nolimits}
\def\sin{\mathop{\rm sin}\nolimits}
\def\cos{\mathop{\rm cos}\nolimits}
\def\var{\mathop{\rm var}}
\def\Re{\mathop{\rm Re}\nolimits}
\def\Sp{\mathop{\rm Sp}\nolimits}
\def\kp{\mathop{\text{\ae}}\nolimits}
\def\bk{{\bf {k}}}
\def\bp{{\bf {p}}}
\def\bq{{\bf {q}}}
\def\lra{\mathop{\longrightarrow}}
\def\Const{\mathop{\rm Const}\nolimits}

\title{The partition function of gauge supersymmetric Ising
model on 3D regular lattice}

\author {S.N. Vergeles\vspace*{4mm}\footnote{{e-mail:vergeles@itp.ac.ru}}}

\affiliation{{Landau Institute for Theoretical Physics, Russian
Academy of Sciences,}\linebreak {Chernogolovka, Moskow region,
142432 Russia} \linebreak \linebreak {and} \linebreak \linebreak
{Moscow Institute of Physics and Technology,} \linebreak {Department
of Theoretical Physics,} \linebreak {Dolgoprudnyj, Moskow region,
Russia}}

\begin{abstract}
The partition function of the gauge system with gauge group $Z_2$
coupled with Majorana fermions is calculated on the regular 3D cubic
lattice
\end{abstract}


\maketitle

\section{Introduction}
\setcounter{equation}{0}

Recently a new approach for calculation the partition function of
the 2D Ising model on the regular lattice has been suggested
\cite{1}. The idea of this approach is as follows. An independent
generator of Clifford algebra in the matrix representation is
assigned to each vertex of the lattice. Thus the number $\mN$ of
generators of the Clifford algebra can not be less than the number
of vertexes of the lattice or the number of degrees of freedom of
the partition function. Some polynomial in these generators
depending on the "temperature" parameter is defined. This polynomial
is the matrix of orthogonal rotation in spinor representation in the
$\mN$-dimensional Euclidean space. The trace of the polynomial is
proportional to partition function of the $2D$ Ising model. This
statement follows from the fact that the considering trace is
represented as a sum over all closed self-intersecting\footnote{The
self-intersections are admissible only orthogonally.} contours
(loops) on the regular planar lattice, and the positive weight
depending on the temperature is assigned to each edge of the loops.

In this paper the outlined approach is extended to three-dimensional
statistical system on on the regular cubic lattice. The sum over
self-intersecting surfaces is calculated. The closed surfaces are
included as well as the surfaces with boundary. Each face of the
surface is assigned the positive factor $\mu^2$, each edge of the
boundary is assigned the factor $\lambda\mu$. The real numbers
$\lambda$ and $\mu$ depend on the temperature. Thus, the total
positive weight corresponds to each closed surface. However, it
turns out that the total weights corresponding to the surfaces with
boundary depend on the boundary configurations, but not on the
surface configurations with the fixed boundary.

The fact that it is impossible to retain only the closed surfaces
and to remove the surfaces with boundary is the important
characteristics of the considered statistical sum. This means that
the considered system does not involve the 3D Ising model. Indeed,
the Ising model on three-dimensional regular lattice is dual to the
thre-dimensional lattice gauge model with the gauge group $Z_2$
\cite{2}. The partition function of the latter is expressed as the
sum over closed self-intersecting surfaces, and each face of the
surface is assigned the positive factor depending on the temperature
at that. So it is naturally to interpret the boundary of the surface
as creation, propagation and annihilation of the fermion pair
coupled with the gauge field. It will be shown in the subsection 2.5
that these fermions are the lattice analog of the Majorana fermions.

Thus, the calculated partition function can be interpreted as the
partition function of the gauge system on the regular
three-dimensional lattice with the gauge group $Z_2$ interecting
with Majorana fermions\footnote {Possibly, it wold be rather to
denominate the calculated partition function as the quantum
transition amplitude from vacuum to vacuum. However, I leave here
for the value the term "partition function" since the signature of
the used metrics is Euclidean.}.

\section{Formulation of the problem}
\setcounter{equation}{0}

\subsection{The Dirac-Clifford algebra in the matrix representation}

The generators of the Clifford algebra $\{\xi_x\}$ satisfy the
following relations
\begin{gather}
\xi_x\xi_y+\xi_y\xi_x=2\delta_{x,\,y}, \quad
x,\,y,\,\ldots=1,\,\ldots\,,I=2K, \label{Z10}
\end{gather}
and they are assumed as hermitian matrixes of the dimension
$2^K\times2^K$. Such dimension is minimally admissible at given
number of generators. The theory of such matrixes in convenient for
physicist form can be found in \cite{3}
for  The concrete matrix representation of the
Clifford algebra generators is no object. Only some their algebraic
properties are needed. It follows from the algebra (\ref{Z10}) that
the trace of any product of the odd number of $\xi$-matrix is equal
to zero, and
\begin{gather}
\tr\xi_x\xi_y=2^K\delta_{x,y}, \quad
\tr\xi_x\xi_y\xi_z\xi_v=2^K\big(\delta_{x,y}\delta_{z,v}-
\delta_{x,z}\delta_{y,v}+\delta_{x,v}\delta_{y,z}\big), \label{Z20}
\end{gather}
and so on. It is evident that by means of permutations and usage of
the equalities $\xi^2_x=1$ any product of $\xi$-matrixes is reduced
either to the number $\pm 1$ (versus the number of permutations) or
to the product of pairwise different $\xi$-matrixes. According to
Eq. (\ref{Z20}) in the last case  the trace of the product is equal
to zero, and in the firs case it is equal to $\pm2^K$.

\subsection{Description of the degrees of freedom of the system}

Let's enumerate the axes the simple three-dimensional cubic lattice
by the numbers $i=1,\,2,\,3$. Each vertex $\mv_{{\bf x}}$ has the
index composed of three natural numbers ${\bf x}\equiv(m,\,n,\,l)$,
and
\begin{gather}
m=1,\,\ldots,\,M, \quad n=1,\,\ldots,\,N,  \quad l=1,\,\ldots,\,L.
\label{Z30}
\end{gather}
For simplicity the numbers $M,\,N,\,L$ are considered even. It is
assumed that the numbers  $(m,\,n,\,l)$ increase uniformly along the
first, the second and the third axis, correspondingly. Define the
base vectors of the lattice: ${\bf e}_1=(1,\,0,\,0)$, ${\bf
e}_2=(0,\,1,\,0)$, ${\bf e}_3=(0,\,0,\,1)$. Thus, the vector ${\bf
e}_i$ is directed onto positive direction of the $i$-th axis.

Further I believe $I=3MNL$.

Divide the totality of the Clifford algebra generators into 3
groups, each of which consisting of $MNL$ generators. Let's
designate the first, the second and the third group of the
generators as $\{\alpha_{{\bf x}}\}$, $\{\beta_{{\bf x}}\}$ and
$\{\gamma_{{\bf x}}\}$, correspondingly. The edge connecting the
adjacent vertexes $\mv_{{\bf x}}$ and $\mv_{{\bf x}+{\bf e}_i}$ is
designated as $\ml_{{\bf x},\,i}=\ml_{m,\,n,\,l;\,i}$. The face with
vertexes $({\bf x},\,{\bf x}+{\bf e}_j,\,{\bf x}+{\bf e}_j+{\bf
e}_k,\, {\bf x}+{\bf e}_k)$ is designated as $\mf_{{\bf x},\,i}$,
$i=1,\,2,\,3$ (here $j\neq k$ and the scalar products ${\bf
e}_j\cdot{\bf e}_i={\bf e}_k\cdot{\bf e}_i=0$).

By definition, the matrixes $\alpha_{{\bf x}}$, $\beta_{{\bf x}}$
and $\gamma_{{\bf x}}$ are related to the edges $\ml_{{\bf x},\,1}$,
$\ml_{{\bf x},\,2}$ and $\ml_{{\bf x},\,3}$, correspondingly. Below
also the notation
\begin{gather}
{\boldsymbol \xi}_{{\bf x}}=\xi^{(i)}_{{\bf x}}=\left(\alpha_{{\bf
x}},\,\beta_{{\bf x}},\,\gamma_{{\bf x}}\right),  \quad i=1,\,2,\,3.
\label{Z33}
\end{gather}
is used.

\subsection{The definition of the partition function}

Relate to each face the unitary matrix of rotation in the spinor
representation:

The face $\mf_{{\bf x},\,1}$ is related with matrix
\begin{gather}
U^{(1)}_{(m,\,n,\,l)}\equiv\left(\lambda+\mu\,\beta_{{\bf
x}}\gamma_{{\bf x}+{\bf e}_2}\right)\left(\lambda+\mu\,\beta_{{\bf
x}+{\bf e}_3}\gamma_{{\bf x}}\right), \label{Z42}
\end{gather}
the face $\mf_{{\bf x},\,2}$ is related with matrix
\begin{gather}
U^{(2)}_{(m,\,n,\,l)}\equiv\left(\lambda+\mu\,\alpha_{{\bf
x}}\gamma_{{\bf x}+{\bf e}_1}\right)\left(\lambda+\mu\,\alpha_{{\bf
x}+{\bf e}_3}\gamma_{{\bf x}}\right), \label{Z44}
\end{gather}
and the face $\mf_{{\bf x},\,3}$ is related with matrix
\begin{gather}
U^{(3)}_{(m,\,n,\,l)}\equiv\left(\lambda+\mu\,\alpha_{{\bf
x}}\beta_{{\bf x}+{\bf e}_1}\right)\left(\lambda+\mu\,\alpha_{{\bf
x}+{\bf e}_2}\beta_{{\bf x}}\right). \label{Z46}
\end{gather}
Here
\begin{gather}
{\bf x}=(m,\,n,\,l),  \quad  \lambda=\cos\frac{\psi}{2}, \quad
\mu=\sin\frac{\psi}{2}. \label{Z48}
\end{gather}

\psfrag{A1}{\rotatebox{0}{\kern0pt\lower0pt\hbox{{$\ve{x}{+}\ve{e}_2{+}\ve{e}_3$}}}}
\psfrag{A2}{\rotatebox{0}{\kern0pt\lower0pt\hbox{{$\ve{x}{+}\ve{e}_3$}}}}
\psfrag{A3}{\rotatebox{0}{\kern0pt\lower0pt\hbox{{$\ve{x}{+}\ve{e}_1{+}\ve{e}_3$}}}}
\psfrag{A4}{\rotatebox{0}{\kern0pt\lower0pt\hbox{{$\ve{x}$}}}}
\psfrag{A5}{\rotatebox{0}{\kern0pt\lower0pt\hbox{{$\ve{x}{+}\ve{e}_1$}}}}
\psfrag{A6}{\rotatebox{0}{\kern0pt\lower0pt\hbox{{$\ve{x}{+}\ve{e}_1{+}\ve{e}_2$}}}}
\psfrag{A7}{\rotatebox{0}{\kern0pt\lower0pt\hbox{{$\ve{x}{+}\ve{e}_2$}}}}
\psfrag{B1}{\rotatebox{30}{\kern0pt\lower0pt\hbox{{$\beta_{\ve{x}{+}\ve{e}_3}$}}}}
\psfrag{B2}{\rotatebox{0}{\kern0pt\lower0pt\hbox{{$\gamma_{\ve{x}}$}}}}
\psfrag{B3}{\rotatebox{0}{\kern0pt\lower0pt\hbox{{$\alpha_{\ve{x}{+}\ve{e}_2}$}}}}
\psfrag{B4}{\rotatebox{30}{\kern0pt\lower0pt\hbox{{$\beta_{\ve{x}{+}\ve{e}_1}$}}}}
\psfrag{B5}{\rotatebox{0}{\kern0pt\lower0pt\hbox{{$\alpha_{\ve{x}}$}}}}
\psfrag{B6}{\rotatebox{30}{\kern0pt\lower0pt\hbox{{$\beta_{\ve{x}}$}}}}
\psfrag{B7}{\rotatebox{0}{\kern0pt\lower0pt\hbox{{$\gamma_{\ve{x}{+}\ve{e}_2}$}}}}
\psfrag{B8}{\rotatebox{0}{\kern0pt\lower0pt\hbox{{$\alpha_{\ve{x}{+}\ve{e}_3}$}}}}
\psfrag{B9}{\rotatebox{0}{\kern0pt\lower0pt\hbox{{$\gamma_{\ve{x}{+}\ve{e}_1}$}}}}
\begin{center}
\includegraphics[scale=0.7]{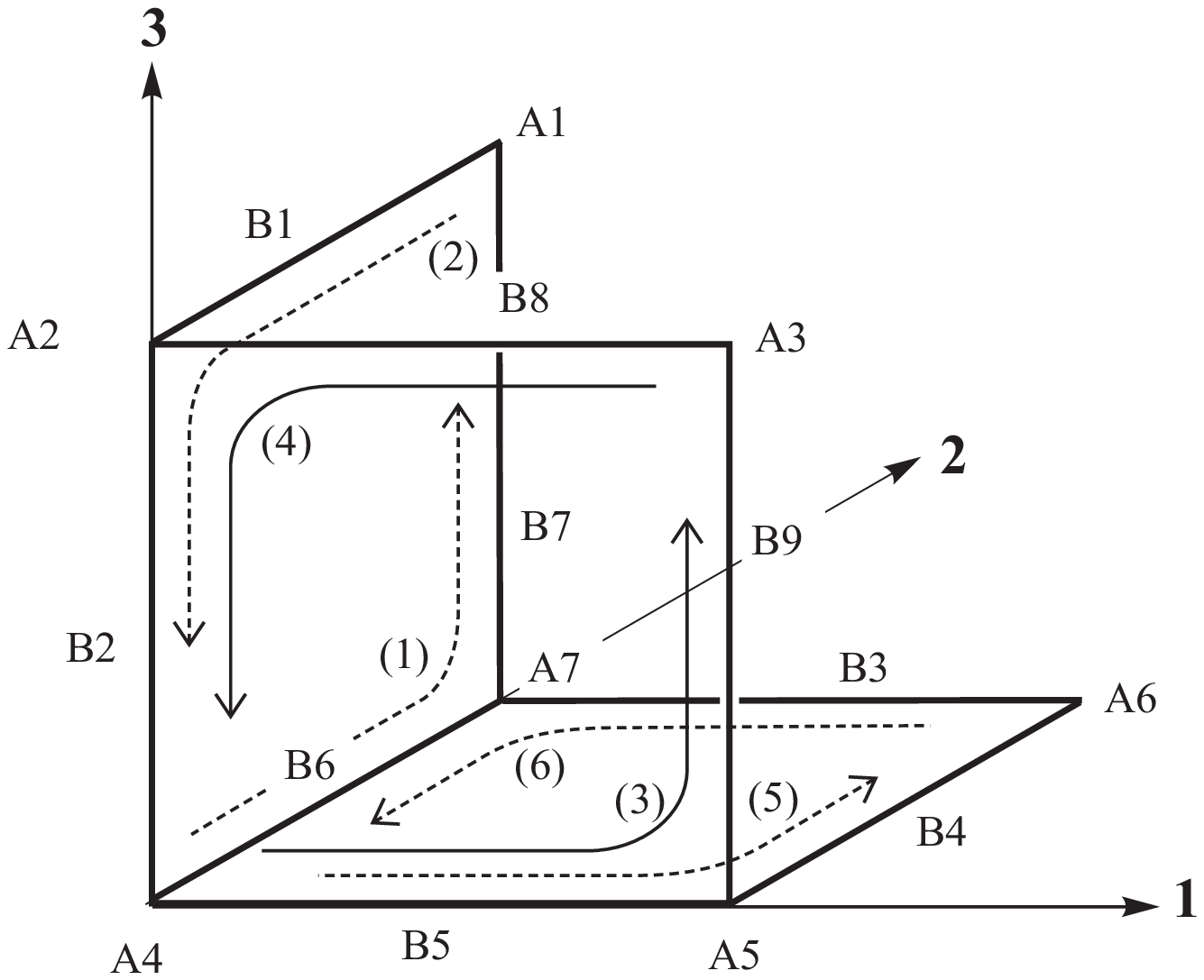}
\end{center}
\vskip8pt

\begin{center}
FIG.~1
\end{center}
\vskip8pt

In fig. 1 a fragment of the lattice containing the vertex ${\bf x}$
and three adjacent faces $\mf_{{\bf x},\,1}$, $\mf_{{\bf x},\,2}$,
$\mf_{{\bf x},\,3}$ is represented, and also the $\xi$-matrixes
related to the corresponding edges are designated. The arrows,
curved orthogonally, lying in the faces and showing the crossing
from one edge to the neighboring edge, represent the rotation
matrixes corresponding to one of the round brackets in the right
hand side of Eqs. (\ref{Z42}), (\ref{Z44}) and (\ref{Z46}); the
directions of the arrows indicate the ordering of $\xi$-matrixes.
For example, the arrows (1) and (2) represent the first and the
second round brackets in Eq. (\ref{Z42}), correspondingly, etc.
Since the round brackets in the right hand sides of Eqs.
(\ref{Z42}), (\ref{Z44}) and (\ref{Z46}) do not commute, so their
order is significant. Below we need the representation of the
quantities (\ref{Z42})-(\ref{Z46}) in the form of the following sums
\begin{gather}
U^{(1)}_{(m,\,n,\,l)}=\left(\lambda^2+\mu^2\beta_{{\bf
x}}\gamma_{{\bf x}+{\bf e}_2}\beta_{{\bf x}+{\bf e}_3}\gamma_{{\bf
x}}\right)+t\lambda\mu\left(\beta_{{\bf x}}\gamma_{{\bf x}+{\bf
e}_2}+\beta_{{\bf x}+{\bf e}_3}\gamma_{{\bf x}}\right), \label{Z50}
\end{gather}
\begin{gather}
U^{(2)}_{(m,\,n,\,l)}=\left(\lambda^2+\mu^2\alpha_{{\bf
x}}\gamma_{{\bf x}+{\bf e}_1}\alpha_{{\bf x}+{\bf e}_3}\gamma_{{\bf
x}}\right)+t\lambda\mu\left(\alpha_{{\bf x}}\gamma_{{\bf x}+{\bf
e}_1}+\alpha_{{\bf x}+{\bf e}_3}\gamma_{{\bf x}}\right), \label{Z51}
\end{gather}
\begin{gather}
U^{(3)}_{(m,\,n,\,l)}=\left(\lambda^2+\mu^2\alpha_{{\bf
x}}\beta_{{\bf x}+{\bf e}_1}\alpha_{{\bf x}+{\bf e}_2}\beta_{{\bf
x}}\right)+t\lambda\mu\left(\alpha_{{\bf x}}\beta_{{\bf x}+{\bf
e}_1}+\alpha_{{\bf x}+{\bf e}_2}\beta_{{\bf x}}\right). \label{Z52}
\end{gather}
In the considered theory
\begin{gather}
t=1. \label{Z53}
\end{gather}
In the case $t\neq1$ the matrixes (\ref{Z50})-(\ref{Z52}) are not
the rotation matrixes and therefore the used here calculation method
turns out useless.

The partition function of the system is proportional to the
following trace of the ordered product of rotation matrixes in
spinor representation
\begin{gather}
Z=C\tr{\cal U}, \label{Z60}
\end{gather}
\begin{gather}
{\cal U}=\left(\hat{P}\prod_l\right){\cal U}_{(l)}, \label{Z70}
\end{gather}
\begin{gather}
{\cal U}_{(l)}=\left(\hat{P}\prod_n\right){\cal U}_{(n,\,l)},
\label{Z80}
\end{gather}
\begin{gather}
{\cal U}_{(n,\,l)}=\left(\hat{P}\prod_m\right)U_{(m,\,n,\,l)}=\ldots
U_{(m-1,\,n,\,l)}U_{(m,\,n,\,l)}U_{(m+1,\,n,\,l)}\ldots. \label{Z90}
\end{gather}
\begin{gather}
U_{(m,\,n,\,l)}\equiv
U^{(1)}_{(m,\,n,\,l)}U^{(2)}_{(m,\,n,\,l)}U^{(3)}_{(m,\,n,\,l)},
\label{Z100}
\end{gather}
Here $C$ is some numerical constant which is not interesting here.
The symbol $\big(\hat{P}\prod_m\big)$ in (\ref{Z90}) denotes the
ordered product of the matrices (\ref{Z100}) over all meaning of $m$
at fixed values of the numbers $(n,\,l)$, so that in this product
the matrix $U_{(m,\,n,\,l)}$ is placed to the left of the matrix
$U_{(m',\,n,\,l)}$ for $m<m'$. The ordered products in (\ref{Z80})
(at fixed $(l)$) and in (\ref{Z70}) are determined analogously.

\subsection{The closed surfaces with self-intersections}

Let's consider the contribution of zeroth power in the parameter $t$
to the matrix (\ref{Z70}). According to Eqs. (\ref{Z50})-(\ref{Z52})
in this case each face gives two contributions to ${\cal U}$: the
first contribution is proportional to the number $\lambda^2$ and the
second one is proportional to the matrix $\{\mu^2\times$(the ordered
product of four $\xi$-matrixes corresponding to four edges of the
face)$\}$. It is obvious that under the trace operator in
(\ref{Z60}) some summands of the matrix ${\cal U}$ are inessential.
For example, according to (\ref{Z20})
\begin{gather}
\tr C\left\{(\lambda^2)^{(3MNL-1)}\mu^2\beta_{{\bf x}}\gamma_{{\bf
x}+{\bf e}_2}\beta_{{\bf x}+{\bf e}_3}\gamma_{{\bf x}}\right\}=0.
\label{S10}
\end{gather}
Only that summands of the matrix ${\cal U}$ give the nonzero
contribution to the partition function in which each contained
$\xi$-matrix is in even power. It follows from here that the
considered part of partition function is represented as the sum over
closed surfaces on the lattice with self-intersections. Each edge of
the surface can belong to two or four faces of the surface. In the
last case the intersection or self-intersection of the surface occur
in the edge. To each face $\mf_{{\bf x},\,i}$ of such surface the
summand from $U^{(i)}_{(m,\,n,\,l)}$, $i=1,\,2,\,3$, proportional to
$\mu^2$ and the fourth power of $\xi$-matrix (see
(\ref{Z50})-(\ref{Z52})), is assigned. The order of the $\xi$-matrix
arrangement in the summand in ${\cal U}$ corresponding to the closed
surface (not obvious connected) is determined by the formulae
(\ref{Z60})-(\ref{Z100}). It follows from the aforesaid that in the
summand in ${\cal U}$ corresponding to any closed surface each
$\xi$-matrix can be contained either in the power 0 (the
corresponding edge does not belong to the surface) either in the
power 2 (the corresponding edge belong to two faces of the surface
only) or in the power 4 (the corresponding edge belong to four faces
of the surface). If $S$ is the number of faces of a the closed
surface, then the contribution to the partition function of the
surface is equal to (at macroscopic dimensions of the lattice the
total number of faces of the lattice can be considered equal to
$3MNL$)
\begin{gather}
\Delta
Z=C\left[2^{3MNL/2}(\lambda^2)^{3MNL}\right]\left(\mu/\lambda\right)^{2S},
\quad  C>0.
\label{Z110}
\end{gather}
Up to the sign the equality (\ref{Z110}) is evident. It remains to
prove only that each closed surface gives the {\it positive}
contribution to the partition function.

Consider the contribution to ${\cal U}$ from the simplest closed
surface without self-intersections: the elementary cube with the
faces $\mf_{{\bf x},\,1}$, $\mf_{{\bf x},\,2}$, $\mf_{{\bf x},\,3}$,
$\mf_{{\bf x}+{\bf e}_1,\,1}$, $\mf_{{\bf x}+{\bf e}_2,\,2}$,
$\mf_{{\bf x}+{\bf e}_3,\,3}$. Here the order of the face
enumeration corresponds to the order of construction from elementary
"bricks" the corresponding summand in ${\cal U}$. Thus
\begin{gather}
\Delta{\cal U}=\left(\beta_{{\bf x}}\gamma_{{\bf x}+{\bf
e}_2}\beta_{{\bf x}+{\bf e}_3}\gamma_{{\bf
x}}\right)\left(\alpha_{{\bf x}}\gamma_{{\bf x}+{\bf
e}_1}\alpha_{{\bf x}+{\bf e}_3}\gamma_{{\bf x}}\right)
\left(\alpha_{{\bf x}}\beta_{{\bf x}+{\bf e}_1}\alpha_{{\bf x}+{\bf
e}_2}\beta_{{\bf x}}\right)\times
\nonumber \\[8pt]
\times\left(\beta_{{\bf x}+{\bf e}_1}\gamma_{{\bf x}+{\bf e}_1+{\bf
e}_2}\beta_{{\bf x}+{\bf e}_1+{\bf e}_3}\gamma_{{\bf x}+{\bf
e}_1}\right)\left(\alpha_{{\bf x}+{\bf e}_2}\gamma_{{\bf x}+{\bf
e}_1+{\bf e}_2}\alpha_{{\bf x}+{\bf e}_2+{\bf e}_3}\gamma_{{\bf
x}+{\bf e}_2}\right)\times
\nonumber \\[8pt]
\times\left(\alpha_{{\bf x}+{\bf e}_3}\beta_{{\bf x}+{\bf e}_1+{\bf
e}_3}\alpha_{{\bf x}+{\bf e}_2+{\bf e}_3}\beta_{{\bf x}+{\bf
e}_3}\right)=1. \label{Z120}
\end{gather}
In this product each $\xi$-matrix is contained twice.

To prove the positivity of the expression (\ref{Z110}) in general
case let's cut any closed surface by the plane parallel to the plane
$(12)$, intersecting the middles of the edges $\ml_{m,\,n,\,l;\,3}$
with fixed $l$ and $m=1,\,\ldots\,,\,M$, $n=1,\,\ldots\,,\,N$.
Denote this plane by $\mP_l$. In each plane $\mP_l$ there is the
natural structure of the square plane lattice with the vertexes
designated as $\mv_{m,\,n}$ which belong to the  middles of the
edges $\ml_{m,\,n,\,l;\,3}$. The edges of this lattice are
designated by $\ml_{m,\,n;\,1}$ and $\ml_{m,\,n;\,2}$. The edge
$\ml_{m,\,n;\,1}$ connects the vertexes $\mv_{m,\,n}$ and
$\mv_{m+1,\,n}$, and the edge $\ml_{m,\,n;\,2}$ connects the
vertexes $\mv_{m,\,n}$ and $\mv_{m,\,n+1}$. Thus the edge
$\ml_{m,\,n;\,1}$ belong to the face $\mf_{{\bf x},\,2}$ and the
edge $\ml_{m,\,n;\,2}$ belong to the face $\mf_{{\bf x},\,1}$. The
outlined lattice also is denoted as $\mP_l$. It is convenient to
consider that the matrixes $\gamma_{m,\,n,\,l}$ with fixed $l$ and
$m=1,\,\ldots\,,\,M$, $n=1,\,\ldots\,,\,N$ are related to the
vertexes $\mv_{m,\,n}$ of the lattice $\mP_l$.

Let's fix $l$ and represent by thick segments that edges of the
lattice $\mP_l$ which belong to the faces of the considered closed
surface. It is evident that the totality of thick segments or edges
on the lattice $\mP_l$ forms a closed contour with
self-intersections on the cubic plane lattice, so that in one vertex
$\mv_{m,\,n}$ 0, 2 or 4 thick edges can converge. It is known that
the sum over such closed contours with a certain positive weight
assigned to each edge is proportional to the partition function of
2D Ising model (\cite{1}, \cite{4}). Let us transform the matrix
factors corresponding to the faces $\mf_{{\bf x},\,1}$ and
$\mf_{{\bf x},\,2}$ (belonging to the closed surface and
intersecting with the plane $\mP_l$) as follows:
\begin{gather}
\left(\beta_{{\bf x}}\gamma_{{\bf x}+{\bf e}_2}\beta_{{\bf x}+{\bf
e}_3}\gamma_{{\bf x}}\right)=\left(\gamma_{{\bf x}}\gamma_{{\bf
x}+{\bf e}_2}\right)\left(\beta_{{\bf x}}\beta_{{\bf x}+{\bf
e}_3}\right), \label{Z120}
\end{gather}
\begin{gather}
\left(\alpha_{{\bf x}}\gamma_{{\bf x}+{\bf e}_1}\alpha_{{\bf x}+{\bf
e}_3}\gamma_{{\bf x}}\right)=\left(\gamma_{{\bf x}}\gamma_{{\bf
x}+{\bf e}_1}\right)\left(\alpha_{{\bf x}}\alpha_{{\bf x}+{\bf
e}_3}\right). \label{Z130}
\end{gather}
Note that the matrixes $\left(\gamma_{{\bf x}}\gamma_{{\bf x}+{\bf
e}_i}\right)$, $i=1,\,2$ commute with all matrixes $\alpha_{{\bf
x}'}$, $\beta_{{\bf x}'}$ without restrictions and also with the
matrixes $\left(\gamma_{{\bf x}'}\gamma_{{\bf x}'+{\bf
e}_j}\right)$, $j=1,\,2$ for $l\neq l'$. But the matrixes
$\left(\gamma_{{\bf x}}\gamma_{{\bf x}+{\bf e}_i}\right)$, $i=1,\,2$
and $\left(\gamma_{{\bf x}'}\gamma_{{\bf x}'+{\bf e}_j}\right)$,
$j=1,\,2$, for $l=l'$, generally speaking, do not commute since in
the set of four matrixes $\left(\gamma_{{\bf x}},\,\gamma_{{\bf
x}+{\bf e}_i},\,\gamma_{{\bf x}'},\,\gamma_{{\bf x}'+{\bf
e}_j}\right)$ can be found coinciding matrixes. It is seen from here
and from Eqs. (\ref{Z120}), (\ref{Z130}) that in the summand in
${\cal U}$ corresponding to a closed surface one can remove to the
left all matrix factors $\left(\gamma_{{\bf x}}\gamma_{{\bf x}+{\bf
e}_2}\right)$ and $\left(\gamma_{{\bf x}}\gamma_{{\bf x}+{\bf
e}_1}\right)$ without changing the relative arrangement of the
factors themselves. As a result the sign of the summand does not
change. Moreover, the product of all removed to the left
$\gamma$-matrixes is converted into 1. Indeed, the ordering product
of the removed to the left $\gamma$-matrixes (at fixed $l$)
corresponds to the ordering product of $\gamma$-matrixes in the case
of $2D$ Ising model \cite{1} . It was proved that in the last case
the considered product of $\gamma$-matrixes is equal to 1.

Analogously the $\gamma$-matrixes for all others $l$ are mutually
cancelled. Thus in any summand in the matrix ${\cal U}$ we can cut
out all $\gamma$-matrixes without changing the sign.

Further we work analogously. Let's denote the plane cutting a closed
surface through the middles of the edges $\ml_{m,\,n,\,l;\,2}$,
$m=1,\,\,\ldots,\,M$, $l=1,\,\ldots,\,L$ by $\mP_n$. On the plane
$\mP_n$ the natural structure of the right square plane lattice is
present. The vertexes of this lattice (at fixed $n$) are designated
as $\mv_{m,\,l}$. The matrixes $\beta_{m,\,n,\,l}$ are related to
the vertexes $\mv_{m,\,l}$. Taking into account Eqs. (\ref{Z120})
and
\begin{gather}
\left(\alpha_{{\bf x}}\beta_{{\bf x}+{\bf e}_1}\alpha_{{\bf x}+{\bf
e}_2}\beta_{{\bf x}}\right)=\left(\beta_{{\bf x}}\beta_{{\bf x}+{\bf
e}_1}\right)\left(\alpha_{{\bf x}}\alpha_{{\bf x}+{\bf e}_2}\right),
\label{Z140}
\end{gather}
we can (at fixed $n$) remove from the rest of the $\xi$-matrix
product (corresponding to the closed surface) all factors
$\left(\beta_{{\bf x}}\beta_{{\bf x}+{\bf e}_3}\right)$ and
$\left(\beta_{{\bf x}}\beta_{{\bf x}+{\bf e}_1}\right)$ to the left
without changing the relative arrangement of the factors themselves.
As above, the total sign of the $\xi$-matrix product does not
change, and the product of $\beta$-matrixes at each value of $n$ is
converted to 1. Further only the product of $\alpha$-matrixes
remains which is converted to 1 at each fixed $m$.

Thus it is proved that the product of $\xi$-matrixes in each matrix
summand in ${\cal U}$ corresponding to any closed surface is
converted to 1, and the formula (\ref{Z110}) is true.

\subsection{The surfaces with boundary}

Let us consider the contributions to the partition function
(\ref{Z60}) which are nonzero relative to the parameter $t$ in
(\ref{Z50})-(\ref{Z52}).

The elementary example of the surfaces with boundary is represented
in fig. 2. The totality of curved orthogonally thick
arrows\footnote{These arrows are the same as in fig. 1.} forms the
boundary of the surface. The surface itself is the shaded face in
fig. 2 with the vertexes $\mv_2,\,\mv_3,\,\mv_4,\,\mv_6$.

\psfrag{v1}{\rotatebox{0}{\kern0pt\lower0pt\hbox{{$\mv_1$}}}}
\psfrag{v2}{\rotatebox{0}{\kern0pt\lower0pt\hbox{{$\mv_2$}}}}
\psfrag{v3}{\rotatebox{0}{\kern0pt\lower0pt\hbox{{$\mv_3$}}}}
\psfrag{v4}{\rotatebox{0}{\kern0pt\lower0pt\hbox{{$\mv_4$}}}}
\psfrag{v5}{\rotatebox{0}{\kern0pt\lower0pt\hbox{{$\mv_5$}}}}
\psfrag{v6}{\rotatebox{0}{\kern0pt\lower0pt\hbox{{$\mv_6$}}}}
\psfrag{v7}{\rotatebox{0}{\kern0pt\lower0pt\hbox{{$\mv_7$}}}}
\psfrag{v8}{\rotatebox{0}{\kern0pt\lower0pt\hbox{{$\mv_8$}}}}
\psfrag{v9}{\rotatebox{0}{\kern0pt\lower0pt\hbox{{$\mv_9$}}}}
\psfrag{al1}{\rotatebox{0}{\kern0pt\lower0pt\hbox{{$\alpha_1$}}}}
\psfrag{al2}{\rotatebox{0}{\kern0pt\lower0pt\hbox{{$\alpha_2$}}}}
\psfrag{al3}{\rotatebox{0}{\kern0pt\lower0pt\hbox{{$\alpha_3$}}}}
\psfrag{al4}{\rotatebox{0}{\kern0pt\lower0pt\hbox{{$\alpha_4$}}}}
\psfrag{be1}{\rotatebox{0}{\kern0pt\lower0pt\hbox{{$\beta_1$}}}}
\psfrag{be2}{\rotatebox{0}{\kern0pt\lower0pt\hbox{{$\beta_2$}}}}
\psfrag{be3}{\rotatebox{0}{\kern0pt\lower0pt\hbox{{$\beta_3$}}}}
\psfrag{ga1}{\rotatebox{0}{\kern0pt\lower0pt\hbox{{$\gamma_1$}}}}
\psfrag{ga2}{\rotatebox{0}{\kern0pt\lower0pt\hbox{{$\gamma_2$}}}}
\psfrag{ga3}{\rotatebox{0}{\kern0pt\lower0pt\hbox{{$\gamma_3$}}}}
\psfrag{ga4}{\rotatebox{0}{\kern0pt\lower0pt\hbox{{$\gamma_4$}}}}
\begin{center}
\includegraphics[scale=0.7]{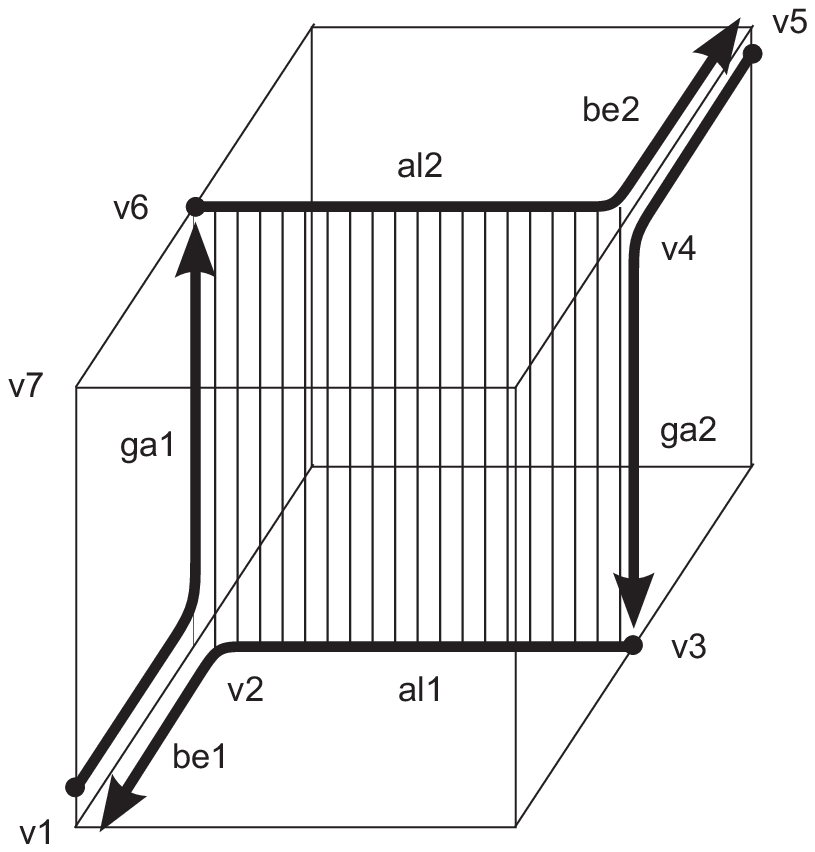}
\end{center}
\vskip8pt

\begin{center}
FIG. 2
\end{center}
\vskip8pt

The contribution to the partition function of the surface with
boundary represented in fig. 2 is proportional to $t^4$. According
to the rule the contribution of this surface to ${\cal U}$ is
proportional to the expression
\begin{gather}
\Delta_1{\cal U}\sim(\beta_1\gamma_1)
(\alpha_1\beta_1)(\alpha_1\gamma_2\alpha_2\gamma_1)(\beta_2\gamma_2)(\alpha_2\beta_2)=-1.
\label{B10}
\end{gather}
Here and below the proportionality coefficients are positive. In
(\ref{B10}) the factors $(\beta_1\gamma_1)$ contained in parentheses
are the contributions from the curved arrow in the face with
vertexes $\mv_1,\,\mv_2,\,\mv_6,\,\mv_7$. Generally, the set of
factors contained in parentheses is the factor in $\Delta{\cal U}$
either from one arrow in some face (only two $\xi$-matrixes), or the
factor from the whole face (four $\xi$-matrixes). Each face gives no
more than one factor which can be either zeroth power relative to
the $\xi$-matrixes and proportional to $\lambda^2$, either the
second power (represented by one curved arrow and proportional to
$t$), or the fourth power relative to $\xi$-matrixes. For example,
the shaded face with vertexes $\mv_2,\,\mv_3,\,\mv_4,\,\mv_6$ gives
the contribution of the fourth power relative to $\xi$-matrixes:
$(\alpha_1\gamma_2\alpha_2\gamma_1)$.

Now let's consider the contribution to ${\cal U}$ from the surface
with boundary represented in fig. 3. As in the previous example, the
curved orthogonally thick arrows form the boundary of the surface,
and the surface itself consists of the three shaded faces. The
contribution of this surface to the matrix ${\cal U}$ is
proportional to $t^6$ and the matrix
\begin{gather}
\Delta_2{\cal U}\sim(\beta_1\gamma_1)(\alpha_1\beta_1)
(\alpha_1\gamma_2\alpha_2\gamma_1)(\beta_3\gamma_2)(\beta_2\gamma_3)\times
\nonumber \\[8pt]
\times(\alpha_2\beta_3\alpha_3\beta_2)(\alpha_3\gamma_4\alpha_4\gamma_3)
(\beta_4\gamma_4)(\alpha_4\beta_4)=1. \label{B20}
\end{gather}

\psfrag{f301}{\rotatebox{0}{\kern0pt\lower0pt\hbox{{$\mv_{10}$}}}}
\psfrag{be4}{\rotatebox{0}{\kern0pt\lower0pt\hbox{{$\beta_4$}}}}
\psfrag{ga1}{\rotatebox{0}{\kern0pt\lower0pt\hbox{{$\gamma_1$}}}}
\begin{center}
\includegraphics[scale=0.7]{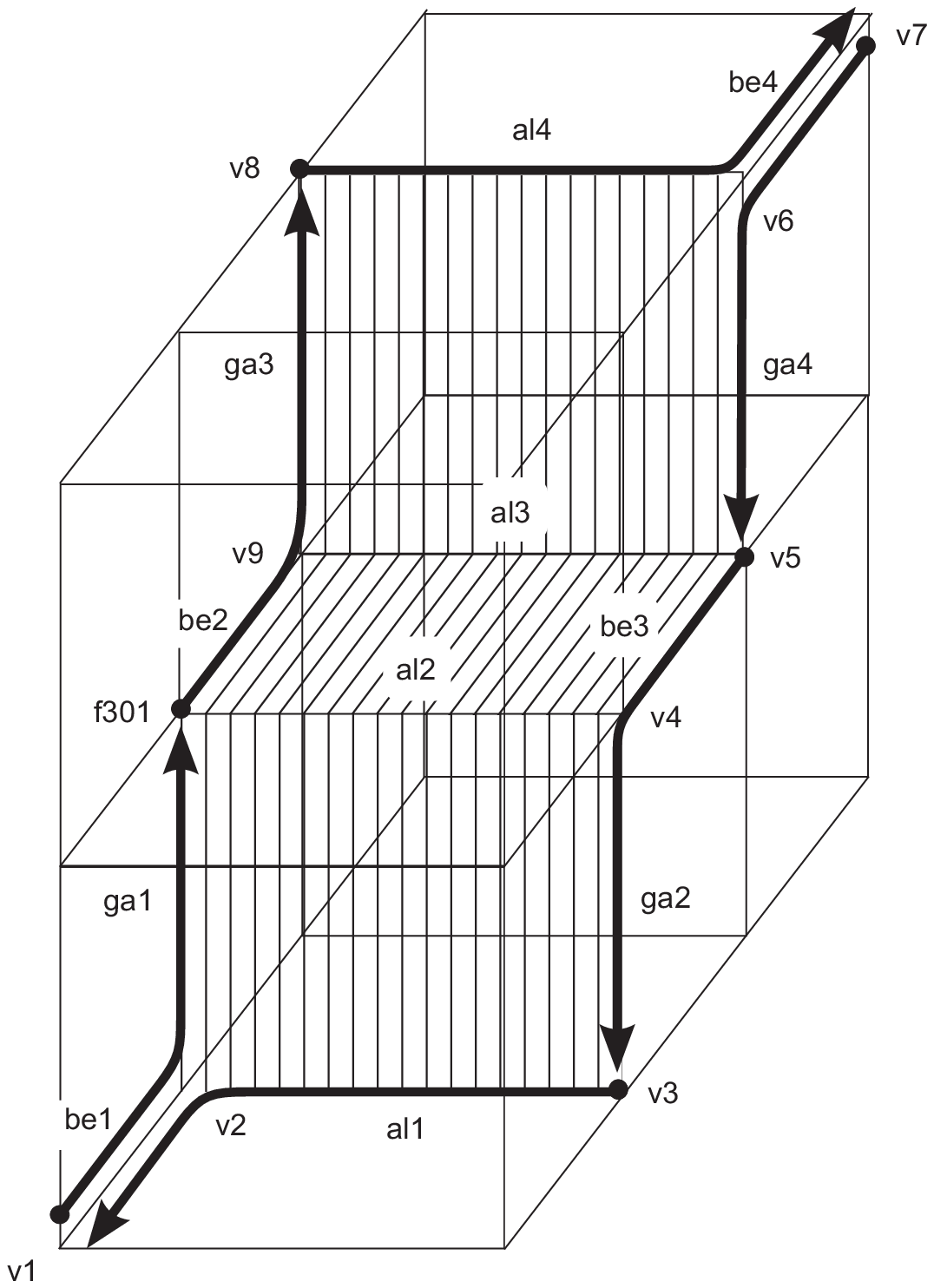}
\end{center}
\vskip8pt

\begin{center}
FIG. 3
\end{center}
\vskip8pt

It is seen from the comparison of the right hand sides of the
relations (\ref{B10}) and (\ref{B20}) that the factor in the summand
of the partition function corresponding to a surface with boundary
can be of any sign. The more detailed consideration of this problem
shows that the signs of the factors of the surfaces with boundary
depend on the boundary configuration but not on the surface
configuration with given boundary.

The outlined situation resembles the situation in the case of
Majorana fermions on three-dimensional lattice coupled with Abelian
gauge field. It is shown in Appendix B that in the last case the
fermion contribution to the partition function also is represented
as a sum over the surfaces with boundary, and the boundary of the
surface is interpreted as creation, propagation and annihilation of
the fermion pair. As in the studied here model, each such surface
gives some factor in one of summand of all functional integral. The
sign of this factor can be either positive or negative, and the sign
is determined by the boundary configuration but not by the surface
configuration with given boundary. The modulo of this factor is
proportional to $\epsilon^s$, where $\epsilon$ is some positive
number and $s$ is the number of faces of the corresponding surface.

Stated above founds to accept the hypotheses that the studied here
partition function is the partition function of the abelian  gauge
system with the gauge group $Z_2$ (this is evident) coupled with
Majorana fermions. Just owing to this hypotheses the word
"supersymmetric" is introduced into the title of the paper.

In connection with this hypotheses it is necessary to make
the following comment.

As it is known \cite{5}, it is impossible to construct on the periodic lattice
a satisfactory variant of the local action of Dirac fermions which transforms
into the well known Dirac action in the continual limit.
In particular, the problem consists in the impossibility of
construction of fermion action on the periodic lattices describing
{\it only one chiral (Weyl)} fermion field and transforming into
the usual action of chiral field in the continual limit. This
problem is known as a "fermion doubling" problem. It is known
also that at the cost of increase the lattice dimension
one can construct only one Weyl field on the lattice hypersurface
of the codimension one. For example, on the 4+1-dimensional
lattice one can construct the action of only one Weyl field
on 4-dimensional sublattice \cite{6}. But this construction
does not seem refined and therefore prospective. Besides the problem of
coupling of the such Weyl fields with the gauge field is
not solved in general case beyond the framework of perturbation theory.

From here one can make the following conclusion: the approach to the
study of the lattice gauge systems including fermion degrees of
freedom which is based on the usage of a lattice action is incorrect
in general. But another approach can be used for the study the
gauge-fermion systems on the lattice. Indeed, the exponent action
summed (integrated) over the degrees of freedom of the system is
necessary only for finding the quantum transition amplitudes. Though
in the continuous theory the method of finding of transition
amplitudes with the help of functional integral of the exponent
action is very effective, we see that in the lattice theories on the
such way can arise serious difficulties by reason of lack of the
action itself. Fundamentally this obstacle can be overcame by means
of the direct definition of transition amplitudes omitting the stage
of action construction\footnote{For example, the specification of
the diagram technique rules defines the elementary transition
amplitudes and the rules of their superposition and construction of
any amplitudes from the elementary amplitudes. Thus the knowledge of
the diagram technique rules free from the necessity of the action
knowledge and the study of functional integral leading to these
rules.}. This approach seems more perspective when studying the
fundamental phenomena, especially in the case of discrete
space-time: it is not ruled out that exactly quantum transition
amplitudes are the most elementary mathematical objects in
fundamental physics. Such approach is realized in this work.

According to the aforesaid the question of the representation of
the partition function (quantum vacuum-vacuum transition amplitude)
(\ref{Z60}) in the form of exponent action
summed over the degrees of freedom of the system is not correct entirely.
Indeed, the considering sum over closed surfaces is nothing but
the partition function of the gauge Ising model, that is summed over
the system degrees of freedom exponent lattice action of the gauge field
without matter with the gauge group $Z_2$. But the sum over
surfaces with boundary unlikely can be expressed in a similar manner
as an integral over fermions of the corresponding exponent,
though this sum must be interpreted as a Majorana fermions contribution.
The last statement follows from the comparison of the
deciding properties of the weights of the surfaces with boundary
(the signs of the weights depend on the boundary surface shape only)
in the considered model and in the case of the lattice Majorana fermions
(see Appendix B). But the exact coincidence of these weights unlikely
can be achieved by means of the search of the appropriate
lattice fermion model because the fermion doubling problem prevents.
The correct problem statement seems to be as follows: Suppose
that in the model presented by the quantum transition amplitudes
(but not by the action) the second kind phase transition exists,
that is the long-wave or macroscopic limit takes place.
Then one must study the properties of the amplitudes in the
long-wave limit and compare them with the properties of amplitudes
in the continuous theories . As a result the corresponding
identification can be determined in the phase transition point.

Note that the free energy of the studied model
has a peculiarity by the parameter $\psi$ (see below).
This means that there is the phase transition in the model.
But the nature of this phase transition is not studied here.

\section{The first stage of diagonalization of the matrix ${\cal O}_{x,\,y}$}
\setcounter{equation}{0}

In the Appendix A the formulas are given which express the partition
function (\ref{Z60}) through the eigenvalues of the orthogonal
rotation matrix ${\cal O}_{x,\,y}$ (see
(\ref{repr1})-(\ref{repr4})). Therefore here and in the subsequent
section the problem of finding the matrix ${\cal O}_{x,\,y}$
eigenvalues is solved and the solution is given in (\ref{eig180}).
The reader wishing to omit these bulky calculations can continue the
reading of the paper starting from the Eqs. (\ref{eig180}).

In statistical limit,  when $M,\,N,\,L\rightarrow\infty$, the
problem of matrix ${\cal O}_{{\bf x},{\bf y}}$ diagonalization
simplifies radically since the matrices ${\cal O}_{{\bf x},{\bf
x}+{\bf z}}$ and $\omega_{{\bf x},{\bf x}+{\bf z}}$ become dependent
only on ${\bf z}$ at relatively small distances from boundary of the
lattice. This property is known as translational invariance.
Therefore the diagonalization of these matrices is performed by
means of Fourier transformation, i.e. by passing to quasi-momentum
representation. The following complete orthonormal set of functions
on the lattice is used for that purpose:
\begin{gather}
|p\rangle\equiv\Psi_p(m)=\dfrac{1}{\sqrt{M}}\,e^{ipm}, \quad
|q\rangle\equiv\Psi_q(n)=\dfrac{1}{\sqrt{N}}\,e^{iqn}, \quad
|r\rangle\equiv\Psi_r(l)=\dfrac{1}{\sqrt{L}}\,e^{irl},
\nonumber \\[8pt]
p=-\dfrac{\pi(M-2)}{M},\,-\dfrac{\pi(M-4)}{M},\,\ldots,\,0,\,\dfrac{2\pi}{M},\,\ldots,\,\pi,
\nonumber \\[8pt]
q=-\dfrac{\pi(N-2)}{N},\,-\dfrac{\pi(N-4)}{N},\,\ldots,\,0,\,\dfrac{2\pi}{N},\,\ldots,\,\pi,
\nonumber \\[8pt]
r=-\dfrac{\pi(L-2)}{L},\,-\dfrac{\pi(L-4)}{L},\,\ldots,\,0,\,\dfrac{2\pi}{L},\,\ldots,\,\pi,
\nonumber \\[10pt]
|{\bf k}\rangle\equiv\Psi_{{\bf k}}({\bf
x})=\Psi_p(m)\Psi_q(n)\Psi_r(l), \quad {\bf k}=(p,\,q,\,l),
\nonumber \\[10pt]
\sum_{{\bf x}}\overline{\Psi}_{{\bf k}}({\bf x})\Psi_{{\bf k}'}({\bf
x})=\delta_{{\bf k}{\bf k}'} \longleftrightarrow \sum_{{\bf
k}}\Psi_{{\bf k}}({\bf x})\overline{\Psi}_{{\bf k}}({\bf x}')=
\delta_{{\bf x}{\bf x}'}.
\label{repr5}
\end{gather}

Let's divide the problem into the successive series of steps.

It follows from Eqs. (\ref{Z70})-(\ref{Z90}) and (\ref{repr2})
\begin{gather}
{\cal U}^{\dag}_{(n,\,l)}\xi_{x}{\cal U}_{(n,\,l)}={\cal
O}^{(n,\,l)}_{x,\,y}\xi_{y}, \label{eig10}
\end{gather}
\begin{gather}
{\cal U}^{\dag}_{(l)}\xi_{x}{\cal U}_{(l)}={\cal
O}^{(l)}_{x,\,y}\xi_{y},  \quad  {\cal O}^{(l)}=\ldots{\cal
O}^{(n-1,\,l)}{\cal O}^{(n,\,l)}{\cal O}^{(n+1,\,l)}\ldots\,,
\label{eig20}
\end{gather}
\begin{gather}
{\cal O}=\ldots{\cal O}^{(l-1)}{\cal O}^{(l)}{\cal
O}^{(l+1)}\ldots\,. \label{eig30}
\end{gather}

\subsection{The first step}

To find the matrixes ${\cal O}^{(n,\,l)}_{x,\,y}$ we use the
elementary equations
\begin{gather}
\left(\lambda+\mu\xi_x\xi_y\right)^{\dag}\xi_x
\left(\lambda+\mu\xi_x\xi_y\right)= (\cos\psi)\xi_x+(\sin\psi)\xi_y,
\label{EV4}
\end{gather}
\begin{gather}
\left(\lambda+\mu\xi_x\xi_y\right)^{\dag}\xi_y
\left(\lambda+\mu\xi_x\xi_y\right)=(\cos\psi)\xi_y-(\sin\psi)\xi_x,
\label{EV5}
\end{gather}
\begin{gather}
\left(\lambda+\mu\xi_x\xi_y\right)^{\dag}\xi_z
\left(\lambda+\mu\xi_x\xi_y\right)=\xi_z, \quad z\neq x,\,y
\label{EV6}
\end{gather}
which follow from (\ref{Z10}) and (\ref{Z48}). In
(\ref{EV4})-(\ref{EV6}) $x\neq y$.

The direct calculation with the help of Eqs. (\ref{Z90}),
(\ref{Z100}) and (\ref{EV4})-(\ref{EV6}) gives (remember that ${\bf
x}=(m,\,n,\,l)$):
\begin{gather}
{\cal U}_{(n,\,l)}^{\dag}\alpha_{{\bf x}}{\cal
U}_{(n,\,l)}=\Big[\left(\cos^2\psi\right)\alpha_{{\bf
x}}+\left(\cos^3\psi\sin\psi\right)\beta_{{\bf x}+{\bf e}_1}+
\nonumber \\[8pt]
+\left(\cos^2\psi\sin\psi\right)\gamma_{{\bf x}+{\bf
e}_1}+\left(-\cos\psi\sin^2\psi\right)\alpha_{{\bf x}+{\bf e}_1+{\bf
e}_3}+\left(-\sin^2\psi\right)\beta_{{\bf x}+{\bf e}_1+{\bf
e}_3}\Big]+
\nonumber \\[8pt]
+\Big[\left(-\cos^2\psi\sin^2\psi\right)\alpha_{{\bf x}+{\bf
e}_1+{\bf e}_2}+\left(\cos\psi\sin^2\psi\right)\gamma_{{\bf x}+{\bf
e}_1+{\bf e}_2}\Big], \label{EV10}
\end{gather}
\begin{gather}
{\cal U}_{(n,\,l)}^{\dag}\beta_{{\bf x}}{\cal
U}_{(n,\,l)}=\Big[\left(-\sin\psi\right)\alpha_{{\bf x}-{\bf
e}_1}+\left(\cos^3\psi\right)\beta_{{\bf x}}\Big]+
\nonumber \\[8pt]
+\Big[\left(-\cos^2\psi\sin\psi\right)\alpha_{{\bf x}+{\bf
e}_2}+\left(\cos\psi\sin\psi\right)\gamma_{{\bf x}+{\bf e}_2}\Big],
\label{EV20}
\end{gather}
\begin{gather}
{\cal U}_{(n,\,l)}^{\dag}\gamma_{{\bf x}}{\cal
U}_{(n,\,l)}=\Big[\left(-\cos\psi\sin\psi\right)\alpha_{{\bf x}-{\bf
e}_1}+\left(-\cos^2\psi\sin^2\psi\right)\beta_{{\bf x}}+
\nonumber \\[8pt]
+\left(\cos^3\psi\right)\gamma_{{\bf
x}}+\left(-\cos^2\psi\sin\psi\right)\alpha_{{\bf x}+{\bf
e}_3}+\left(-\cos\psi\sin\psi\right)\beta_{{\bf x}+{\bf e}_3}\Big]+
\nonumber \\[8pt]
+\Big[\left(\cos\psi\sin^3\psi\right)\alpha_{{\bf x}+{\bf
e}_2}+\left(-\sin^3\psi\right)\gamma_{{\bf x}+{\bf e}_2}\Big],
\label{EV30}
\end{gather}
\begin{gather}
{\cal U}_{(n,\,l)}^{\dag}\alpha_{{\bf x}+{\bf e}_3}{\cal
U}_{(n,\,l)}=\Big[\left(\sin\psi\right)\gamma_{{\bf
x}}+\left(\cos\psi\right)\alpha_{{\bf x}+{\bf e}_3}\Big],
\label{EV40}
\end{gather}
\begin{gather}
{\cal U}_{(n,\,l)}^{\dag}\beta_{{\bf x}+{\bf e}_3}{\cal
U}_{(n,\,l)}=\Big[\left(\cos\psi\sin\psi\right)\gamma_{{\bf
x}}+\left(-\sin^2\psi\right)\alpha_{{\bf x}+{\bf
e}_3}+\left(\cos\psi\right)\beta_{{\bf x}+{\bf e}_3}\Big],
\label{EV50}
\end{gather}
\begin{gather}
{\cal U}_{(n,\,l)}^{\dag}\alpha_{{\bf x}+{\bf e}_2}{\cal
U}_{(n,\,l)}=\Big[\left(\sin\psi\right)\beta_{{\bf
x}}\Big]+\Big[\left(\cos\psi\right)\alpha_{{\bf x}+{\bf e}_2}\Big],
\label{EV60}
\end{gather}
\begin{gather}
{\cal U}_{(n,\,l)}^{\dag}\gamma_{{\bf x}+{\bf e}_2}{\cal
U}_{(n,\,l)}=\Big[\left(-\cos\psi\sin\psi\right)\beta_{{\bf
x}}\Big]+\Big[\left(\sin^2\psi\right)\alpha_{{\bf x}+{\bf
e}_2}+\left(\cos\psi\right)\gamma_{{\bf x}+{\bf e}_2}\Big].
\label{EV70}
\end{gather}
All other $\xi$-matrixes in $\left[{\cal
U}_{(n,\,l)}^{\dag}\ldots{\cal U}_{(n,\,l)}\right]$ remain
unchangeable. Therefore the right-hand sides of Eqs.
(\ref{EV10})-(\ref{EV70}) define completely the matrixes ${\cal
O}^{(n,\,l)}_{x,\,y}$. It is evident that if the vector ${\bf x}$ is
not too close to the boundary of the lattice, then Eqs.
(\ref{EV10})-(\ref{EV70}) are invariant relative to the synchronous
change of the number $m$ in ${\bf x}$ in all summands of these
equations. In the lattice size infinite limit this means the
translational invariance relative to the shifts $m\rightarrow m+m'$
along the first axis. The translational invariance is used for the
partial diagonalization of the matrix ${\cal O}^{(n,\,l)}_{x,\,y}$
by Fourier transformation along the first axis:
\begin{gather}
{\boldsymbol\xi}_{n,\,l}(p)=\sum_m\overline{\Psi}_p(m){\boldsymbol\xi}_{m,\,n,\,l}
={\boldsymbol\xi}^{\dag}_{n,\,l}(-p),
\nonumber \\[8pt]
\left[\xi^{(i)}_{n,\,l}(p),\,\xi^{(j)\,\dag}_{n',\,l'}(p')\right]_+=2\delta_{i,\,j}\delta_{(n,\,l),(n',\,l')}
\delta_{p,\,p'},  \quad   i,\,j=1,\,2,\,3,
\label{EV80}
\end{gather}
\begin{gather}
{\cal U}_{(n,\,l)}^{\dag}\xi^{(i)}_{n',\,l'}(p){\cal U}_{(n,\,l)}=
\sum_{j,\,n'',\,l''}\left[\sum_{m'}{\cal
O}^{(n,\,l)}_{m,\,n',\,l',\,i;\,m+m',n'',\,l'',\,j}e^{ip\,m'}\right]\xi^{(j)}_{n'',\,l''}(p).
\label{EV90}
\end{gather}
The matrix in the square brackets in the right-hand side of Eq.
(\ref{EV90}) is designated $\left[{\cal
O}^{(n,\,l)}(p)\right]_{n',\,l',\,i;\,n'',\,l'',\,j}$ and it is
calculated with the help of Eqs. (\ref{EV10})--(\ref{EV70}). The
essential part of this matrix is written out below. Under the
"essential part" here the part of matrix elements with
$l'=l,\,(l+1)$ and $l''=l,\,(l+1)$ is implied. Indeed, the linear
operator $\left[{\cal
O}^{(n,\,l)}(p)\right]_{n',\,l',\,i;\,n'',\,l'',\,j}$ acts trivially
for the quantities ${\boldsymbol\xi}_{n,\,l'}(p)$ with $l'\neq
l,\,(l+1)$. Therefore it is convenient to order the quantities
${\boldsymbol\xi}_{n,\,l'}(p)$ as follows:
\begin{gather}
\ldots,\,v_{n-1}(p),\,v_n(p),\,v_{n+1}(p),\,\ldots, \label{EV100}
\end{gather}
where $v_n(p)$ are the six-dimensional vectors of the form
\begin{eqnarray}
v_n(p)= \left(
\begin{array}{c}
\alpha_{n,\,l}(p)\\
\beta_{n,\,l}(p)\\
\gamma_{n,\,l}(p)\\
\alpha_{n,\,l+1}(p)\\
\beta_{n,\,l+1}(p)\\
\gamma_{n,\,l+1}(p)
\end{array} \right)
\label{EV110}
\end{eqnarray}
In the such designations the matrixes $\left[{\cal
O}^{(n,\,l)}(p)\right]_{n',\,n''}$ take the form of block-diagonal
matrixex, and each block is the $6\times 6$-matrix, the index $n''$
grows to the right and the index $n'$ grows down.
\begin{eqnarray}
\left[{\cal O}^{(n,\,l)}(p)\right]_{n',\,n''}=
 \left(
\begin{array}{lcccccr}
\cdot & \cdot & \cdot & \cdot & \cdot & \cdot & \cdot \\
\cdot & 1 & 0 & 0 & 0 & 0 & \cdot \\
\cdot & 0 & 1 & 0 & 0 & 0 & \cdot \\
\cdot & 0 & 0 & A(p) & B(p) & 0 & \cdot \\
\cdot & 0 & 0 & D & C & 0 & \cdot \\
\cdot & 0 & 0 & 0 & 0 & 1 & \cdot \\
\cdot & \cdot & \cdot & \cdot & \cdot & \cdot & \cdot
\end{array} \right)
\begin{array}{l}
 \\
 \\
 \\
\scriptstyle{n^\prime = n} \\
\scriptstyle{n^\prime = n+1} \\
\\
\\
\end{array}
. \label{EV120}
\end{eqnarray}
Here $A(p),\,B(p),\,D,\,C$ are $6\times6$-matrixes the same for all
matrixes $\left[{\cal O}^{(n,\,l)}(p)\right]_{n',\,n''}$. The
matrixes $A(p)$ and $C$ are placed on main diagonal on the points
$n$ and $(n+1)$, correspondingly. According to Eqs.
(\ref{EV10})-(\ref{EV70}) and (\ref{EV90})
\begin{eqnarray}
A(p)=\kern250pt
\nonumber \\[8pt]
\scriptsize =\!\left(\!\!\!\!
\begin{array}{cccccr}
\left(\cos^2\psi\right) & \left(e^{ip}\cos^3\psi\sin\psi\right) &
\left(e^{ip}\cos^2\psi\sin\psi\right) &
\left(-e^{ip}\cos\psi\sin^2\psi\right) & \left(-e^{ip}\sin^2\psi\right) & 0  \\
\left(-e^{-ip}\sin\psi\right) & \left(\cos^3\psi\right) & 0 & 0 & 0 & 0  \\
\left(-e^{-ip}\cos\psi\sin\psi\right) &
\left(-\cos^2\psi\sin^2\psi\right) &
\left(\cos^3\psi\right) & \left(-\cos^2\psi\sin\psi\right) & \left(-\cos\psi\sin\psi\right) & 0   \\
0 & 0 & \left(\sin\psi\right) & \left(\cos\psi\right) & 0 & 0   \\
0 & 0 & \left(\cos\psi\sin\psi\right) & \left(-\sin^2\psi\right) & \left(\cos\psi\right) & 0   \\
0 & 0 & 0 & 0 & 0 & 1
\end{array} \!\!\right)\!,
\label{EV130}
\end{eqnarray}
\Large
\begin{eqnarray}
B(p)=  \left(
\begin{array}{cccccr}
\left(-e^{ip}\cos^2\psi\sin^2\psi\right) & 0 &
\left(e^{ip}\cos\psi\sin^2\psi\right) &
0 & 0 & 0  \\
\left(-\cos^2\psi\sin\psi\right) & 0 & \left(\cos\psi\sin\psi\right) & 0 & 0 & 0  \\
\left(\cos\psi\sin^3\psi\right) & 0 & \left(-\sin^3\psi\right) & 0 & 0  & 0   \\
0 & 0 & 0 & 0 & 0 & 0   \\
0 & 0 & 0 & 0 & 0 & 0   \\
0 & 0 & 0 & 0 & 0 & 0
\end{array} \right)=
\nonumber \\[8pt]
 =\left(
\begin{array}{c}
\left(e^{ip}\cos\psi\sin^2\psi\right) \\
\left(\cos\psi\sin\psi\right) \\
\left(-\sin^3\psi\right) \\
0 \\
0 \\
0
\end{array} \right)
\Big(
\begin{array}{cccccr}
\left(-\cos\psi\right) & 0 & 1 & 0 & 0 & 0
\end{array} \Big),
\label{EV140}
\end{eqnarray}
\begin{eqnarray}
C=  \left(
\begin{array}{cccccr}
\left(\cos\psi\right) & 0 & 0 &
0 & 0 & 0  \\
0 & 1 & 0 & 0 & 0 & 0  \\
\left(\sin^2\psi\right) & 0 & \left(\cos\psi\right) & 0 & 0  & 0   \\
0 & 0 & 0 & 1 & 0 & 0   \\
0 & 0 & 0 & 0 & 1 & 0   \\
0 & 0 & 0 & 0 & 0 & 1
\end{array} \right),
\label{EV150}
\end{eqnarray}
\begin{eqnarray}
D=  \left(
\begin{array}{cccccr}
0 & \left(\sin\psi\right) & 0 & 0 & 0 & 0  \\
0 & 0 & 0 & 0 & 0 & 0  \\
0 & \left(-\cos\psi\sin\psi\right) & 0 & 0 & 0  & 0   \\
0 & 0 & 0 & 0 & 0 & 0   \\
0 & 0 & 0 & 0 & 0 & 0   \\
0 & 0 & 0 & 0 & 0 & 0
\end{array} \right).
\label{EV160}
\end{eqnarray}
The matrix (\ref{EV120}) is unitary. This means that
\begin{gather}
A(p)\,[A(p)]^{\dag}+B(p)\,[B(p)]^{\dag}=1,  \quad
C\,C^{\dag}+D\,D^{\dag}=1,
\nonumber \\[8pt]
A(p)\,D^{\dag}+B(p)\,C^{\dag}=0. \label{EV170}
\end{gather}
The direct check shows that the matrixes (\ref{EV130})-(\ref{EV160})
really satisfy the Eqs. (\ref{EV170}).

Thus the first step (finding the matrixes ${\cal O}^{(n,\,l)}(p)$)
is completed.

\subsection{The second step}

To realize the second step in which the matrixes ${\cal O}^{(l)}(p)$
are found we must multiply the matrixes ${\cal O}^{(n,\,l)}(p)$ at
fixed values of the parameters $l$ and $p$ according to the rule in
the right-hand side of Eq. (\ref{eig20}). Thus
\begin{eqnarray}
\left[{\cal O}^{(l)}(p)\right]_{n',\,n''}= \left\{\ldots\left(
\begin{array}{lcccccr}
\cdot & \cdot & \cdot & \cdot & \cdot & \cdot & \cdot \\
\cdot & 1 & 0 & 0 & 0 & 0 & \cdot \\
\cdot & 0 & A(p) & B(p) & 0 & 0 & \cdot \\
\cdot & 0 & D & C & 0 & 0 & \cdot \\
\cdot & 0 & 0 & 0 & 1 & 0 & \cdot \\
\cdot & 0 & 0 & 0 & 0 & 1 & \cdot \\
\cdot & \cdot & \cdot & \cdot & \cdot & \cdot & \cdot
\end{array} \right)\times\right.
\nonumber \\[8pt]
\left.\times \left(
\begin{array}{lcccccr}
\cdot & \cdot & \cdot & \cdot & \cdot & \cdot & \cdot \\
\cdot & 1 & 0 & 0 & 0 & 0 & \cdot \\
\cdot & 0 & 1 & 0 & 0 & 0 & \cdot \\
\cdot & 0 & 0 & A(p) & B(p) & 0 & \cdot \\
\cdot & 0 & 0 & D & C & 0 & \cdot \\
\cdot & 0 & 0 & 0 & 0 & 1 & \cdot \\
\cdot & \cdot & \cdot & \cdot & \cdot & \cdot & \cdot
\end{array} \right) \left(
\begin{array}{lcccccr}
\cdot & \cdot & \cdot & \cdot & \cdot & \cdot & \cdot \\
\cdot & 1 & 0 & 0 & 0 & 0 & \cdot \\
\cdot & 0 & 1 & 0 & 0 & 0 & \cdot \\
\cdot & 0 & 0 & 1 & 0 & 0 & \cdot \\
\cdot & 0 & 0 & 0 & A(p) & B(p) & \cdot \\
\cdot & 0 & 0 & 0 & D & C & \cdot \\
\cdot & \cdot & \cdot & \cdot & \cdot & \cdot & \cdot
\end{array} \right)\ldots\right\}
\begin{array}{l}
 \\
 \\
 \\
\scriptstyle{n^\prime = n} \\
\scriptstyle{n^\prime = n+1} \\
\\
\\
\end{array}
. \label{EV180}
\end{eqnarray}
From here we find by means of the direct and simple calculation
\begin{gather}
\left[{\cal O}^{(l)}(p)\right]_{n,\,n+n'}=\left\{
\begin{array}{cl}
0, &  \mbox{for} \ \ n'<-1 \\[5pt]
D,  & \mbox{for} \ \  n'=-1 \\[5pt]
CA(p), & \mbox{for} \ \  n'=0 \\[5pt]
C\,[B(p)]^{n'}A(p),  & \mbox{for} \ \  n'> 0
\end{array} \right. . \label{EV190}
\end{gather}

It is seen from Eq. (\ref{EV190}) that the matrix $\left[{\cal
O}^{(l)}(p)\right]_{n',\,n''}$ depends only on the difference
$(n''-n')$, i.e. it is translation invariant (at some distance from
the boundary) under the shifts along the second axis \footnote{This
translational invariance follows already from the form of Eq.
(\ref{eig20}).}. This fact enables to perform the further partial
diagonalization of the matrix $\left[{\cal
O}^{(l)}(p)\right]_{n',\,n''}$ by means of Fourier transformation
along the second axis:
\begin{gather}
{\boldsymbol\xi}_{l}(p,\,q)=\sum_n\overline{\Psi}_q(n){\boldsymbol\xi}_{n,\,l}(p)
={\boldsymbol\xi}^{\dag}_{l}(-p,-q),
\nonumber \\[8pt]
\left[\xi^{(i)}_{l}(p,\,q),\,\xi^{(j)\,\dag}_{l'}(p',\,q')\right]_+=2\delta_{i,\,j}\delta_{l,\,l'}
\delta_{p,\,p'}\delta_{q,\,q'},  \quad   i,\,j=1,\,2,\,3.
\label{EV200}
\end{gather}
Now equation (\ref{eig20}) takes the form
\begin{gather}
{\cal U}_{(l)}^{\dag}\xi^{(i)}_{l'}(p,\,q){\cal U}_{(l)}=
\sum_{j,\,l''}\left[\sum_{n'}\left[{\cal
O}^{(l)}(p)\right]_{n,\,l',\,i;\,n+n',\,l'',\,j}e^{iqn'}\right]\xi^{(j)}_{l''}(p,\,q).
\label{EV210}
\end{gather}
In (\ref{EV210}) we returned to the accounting of all values of
indexes $l$. Let's introduce also the designation
\begin{gather}
\sum_{n'}\left[{\cal
O}^{(l)}(p)\right]_{n,\,l',\,i;\,n+n',\,l'',\,j}e^{iqn'}\equiv\left[{\cal
O}^{(l)}(p,\,q)\right]_{l',\,i;\,l'',\,j}. \label{EV220}
\end{gather}
Remind that the matrix (\ref{EV220}) acts trivially for the elements
$\xi^{(j)}_{l''}(p,\,q)$ with $l''\neq l,\,(l+1)$. According to
(\ref{EV140}),
\begin{gather}
\left[B(p)\right]^n=(-\varkappa)^{n-1}B(p),  \quad
\varkappa=\sin^2\psi\left[e^{ip}\cos^2\psi+\sin\psi\right],
\label{EV230}
\end{gather}
Taking into account Eqs. (\ref{EV190}) and (\ref{EV230}) we rewrite
the matrix (\ref{EV220}) in the form
\begin{eqnarray}
{\cal
O}^{(l)}(p,\,q)=\left\{e^{-iq}D+C\left(1+\frac{e^{iq}}{1+e^{iq}\varkappa}B(p)\right)A(p)\right\}=
\left(
\begin{array}{cc}
E(p,\,q)  &  F(p,\,q) \\
H  &  G
\end{array} \right).
\label{EV240}
\end{eqnarray}
Here $E(p,\,q),\,F(p,\,q),\,H,\,G$ are $3\times 3$ matrixes which
are calculated with the help of Eqs. (\ref{EV130})-(\ref{EV160}):
\begin{eqnarray}
E(p,\,q)=\frac{1}{1+e^{iq}\varkappa}\times \kern250pt
\nonumber  \\[8pt]
\tiny \times\left(
\begin{array}{ccc}
\left(\cos^3\psi\right) &
\left(\sin\psi[e^{ip}\cos^2\psi+e^{-iq}+\sin^3\psi]\right) &
\left(e^{ip}\cos^3\psi\sin\psi[e^{iq}\sin\psi+1]\right)  \\
\left(-\sin\psi[e^{iq}\cos^2\psi+e^{i(-p+q)}\sin\psi+e^{-ip}]\right)
& \left(\cos^3\psi\right) &
\left(-e^{iq}\cos^4\psi\sin\psi[e^{ip}\sin\psi-1]\right)   \\
\left(e^{-ip}\cos^2\psi\sin\psi[e^{ip}\sin\psi-1]\right) &
\left(-\cos\psi\sin\psi[e^{-iq}+\sin\psi]\right) &
\left(\cos^2\psi\left[\cos^2\psi+e^{ip}\sin^2\psi\{e^{iq}+\sin\psi\}\right]\right)
\end{array} \right),
\nonumber
\end{eqnarray}
\Large
\begin{eqnarray}
\label{EV250}
\end{eqnarray}
\begin{eqnarray}
F(p,\,q)=\frac{1}{1+e^{iq}\varkappa}\times \kern150pt
\nonumber  \\[8pt]
 \times\left(
\begin{array}{c}
\left(-e^{ip}\cos\psi\sin^2\psi\left[e^{iq}\sin\psi+1\right]\right) \\
\left(e^{iq}\cos^2\psi\sin^2\psi\left[e^{ip}\sin\psi-1\right]\right) \\
\left(-\sin\psi\left[\cos^2\psi+e^{ip}\sin^2\psi\{e^{iq}+\sin\psi\}\right]\right)
\end{array} \right)
\Big(
\begin{array}{ccc}
\left(\cos\psi\right) & 1 & 0
\end{array} \Big),
\label{EV260}
\end{eqnarray}
\begin{eqnarray}
G= \left(
\begin{array}{ccc}
\left(\cos\psi\right) & 0 &  0  \\
\left(-\sin^2\psi\right) & \left(\cos\psi\right) &  0   \\
0 & 0 & 1
\end{array} \right),
\label{EV270}
\end{eqnarray}
\begin{eqnarray}
H= \left(
\begin{array}{ccc}
0 & 0 &  \left(\sin\psi\right)  \\
0 & 0 &  \left(\cos\psi\sin\psi\right)   \\
0 & 0 & 0
\end{array} \right).
\label{EV280}
\end{eqnarray}

The unitarity condition of the matrix (\ref{EV240}) is contained in
the following equations:
\begin{gather}
E(p,\,q)\,[E(p,\,q)]^{\dag}+F(p,\,q)\,[F(p,\,q)]^{\dag}=1,  \quad
G\,G^{\dag}+H\,H^{\dag}=1,
\nonumber \\[8pt]
E(p,\,q)\,H^{\dag}+F(p,\,q)\,G^{\dag}=0. \label{EV290}
\end{gather}
The last equalities are verified directly with the help of Eqs.
(\ref{EV250})-(\ref{EV280}).

Further it is necessary to return to the accounting of all values of
index $l$ and at that it is natural to order the quantities
${\boldsymbol\xi}_{l}(p,\,q)$ as follows (compare with the ordering
in (\ref{EV100}) and (\ref{EV110})):
\begin{gather}
\ldots,\,w_{l-1}(p,\,q),\,w_l(p,\,q),\,w_{l+1}(p,\,q),\,\ldots,
\label{EV350}
\end{gather}
where $w_l(p,\,q)$ are the three-dimensional vectors of the form
\begin{eqnarray}
w_l(p,\,q)= \left(
\begin{array}{c}
\alpha_{l}(p,\,q)\\
\beta_{l}(p,\,q)\\
\gamma_{l}(p,\,q)
\end{array} \right).
\label{EV360}
\end{eqnarray}
In this notations the matrix $\left[{\cal
O}^{(l)}(p,\,q)\right]_{l',\,l''}$ takes the form of a
block-diagonal matrix with the blocks of the size $3\times 3$, and
the index $l''$ grows to the right, the index $l'$ grows down, so
that the matrixes $E(p,\,q)$ and $G$ are placed on the main diagonal
on the places with indexes $l$ and $(l+1)$, correspondingly (compare
with (\ref{EV120})):
\begin{eqnarray}
\left[{\cal O}^{(l)}(p,\,q)\right]_{l',\,l''}=
 \left(
\begin{array}{lcccccr}
\cdot & \cdot & \cdot & \cdot & \cdot & \cdot & \cdot \\
\cdot & 1 & 0 & 0 & 0 & 0 & \cdot \\
\cdot & 0 & 1 & 0 & 0 & 0 & \cdot \\
\cdot & 0 & 0 & E(p,\,q) & F(p,\,q) & 0 & \cdot \\
\cdot & 0 & 0 & H & G & 0 & \cdot \\
\cdot & 0 & 0 & 0 & 0 & 1 & \cdot \\
\cdot & \cdot & \cdot & \cdot & \cdot & \cdot & \cdot
\end{array} \right)
\begin{array}{l}
 \\
 \\
 \\
\scriptstyle{l^\prime = l} \\
\scriptstyle{l^\prime = l+1} \\
\\
\\
\end{array}
. \label{EV400}
\end{eqnarray}

The second step on finding the matrix ${\cal O}^{(l)}(p,\,q)$ is
completed hereon.

\subsection{The third step}

On the third, the last step, the matrix ${\cal O}$ is constructed by means of the ordered
product of the matrixes ${\cal O}^{(l)}$ according to the rule  (\ref{eig30}).
As a result we obtain the matrix invariant relative to the translations
along the third axis. The calculations are identical to that on the second
step, the only difference is in the following substitutions:
\begin{gather}
A(p)\rightarrow E(p,\,q), \quad B(p)\rightarrow F(p,\,q),
\nonumber \\[8pt]
C\rightarrow G,  \quad D\rightarrow H, \label{EV410}
\end{gather}
so that instead of (\ref{EV190}) now we have
\begin{gather}
\left[{\cal O}(p,\,q)\right]_{l,\,l+l'}=\left\{
\begin{array}{cl}
0, &  \mbox{for} \ \ l'<-1 \\[5pt]
H,  & \mbox{for} \ \  l'=-1 \\[5pt]
GE(p,\,q), & \mbox{for} \ \  l'=0 \\[5pt]
G\,[F(p,\,q)]^{l'}E(p,\,q),  & \mbox{for} \ \  l'> 0
\end{array} \right. . \label{EV420}
\end{gather}
Using the translational invariance of the matrix (\ref{EV420}) we pass
to the full Fourier-components of $\xi$-matrixes:
\begin{gather}
{\boldsymbol\xi}(p,\,q,\,r)=\sum_l\overline{\Psi}_r(l){\boldsymbol\xi}_{l}(p,\,q)
={\boldsymbol\xi}^{\dag}(-p,-q,-r),
\nonumber \\[8pt]
\left[\xi^{(i)}(p,\,q,\,r),\,\xi^{(j)\,\dag}(p',\,q',\,r')\right]_+=2\delta_{i,\,j}
\delta_{p,\,p'}\delta_{q,\,q'}\delta_{r,\,r'},  \quad
i,\,j=1,\,2,\,3. \label{EV430}
\end{gather}
\begin{gather}
{\cal U}^{\dag}{\boldsymbol\xi}(p,\,q,\,r)\,{\cal U}=
\left[\sum_{l'}\left[{\cal
O}(p,\,q)\right]_{l,\,l+l'}e^{irl'}\right]{\boldsymbol\xi}(p,\,q,\,r),
\label{EV440}
\end{gather}
\begin{gather}
\sum_{l'}\left[{\cal O}(p,\,q)\right]_{l,\,l+l'}e^{irl'}\equiv {\cal
O}(p,\,q,\,r)=
\nonumber \\[8pt]
=\left\{e^{-ir}H+G\left(1+\frac{e^{ir}}{1+e^{ir}\sigma}F(p,\,q)\right)E(p,\,q)\right\},
\quad
\sigma=\frac{\left(e^{ip}+e^{iq}\right)\cos^2\psi\sin^2\psi}{1+e^{iq}\varkappa}.
\label{EV450}
\end{gather}
Here it was used the fact that according to Eq. (\ref{EV260})
\begin{gather}
[F(p,\,q)]^l=(-\sigma)^{l-1}F(p,\,q). \label{EV455}
\end{gather}

$3\times 3$-matrix ${\cal O}(p,\,q,\,r)$ in (\ref{EV450}) is an unitary matrix:
\begin{gather}
{\cal O}(p,\,q,\,r)\left[{\cal O}(p,\,q,\,r)\right]^{\dag}=1.
\label{EV460}
\end{gather}
The last equality is verified easily with the help of the definition
(\ref{EV450}) and equalities (\ref{EV290}), (\ref{EV455}).

Thus the problem of finding the free energy of the system is reduced to the
visible problem of finding the eigenvalues of $3\times 3$-matrix ${\cal O}(p,\,q,\,r)$ (\ref{EV450}).

\section{The calculation of the eigenvalues of the matrix ${\cal O}(p,\,q,\,r)$}

\setcounter{equation}{0}

Let's give the products of the matrixes in (\ref{EV450}) in the evident form:
\begin{multline}
GE(p,\,q)=\dfrac{1}{1+e^{iq}\varkappa}\times \kern400pt
 \\[20pt]
\scriptsize   \times\left(
\begin{array}{cc}
\left(\cos^4\psi\right) &
\left(\cos\psi\sin\psi[e^{ip}\cos^2\psi+e^{-iq}+\sin^3\psi]\right)   \\ \\
\left(-\cos\psi\sin\psi\left[\cos^2\psi\sin\psi+e^{iq}\cos^2\psi+e^{i(-p+q)}\sin\psi+
e^{-ip}\right]\right) &
\left(\cos^4\psi-\sin^3\psi[e^{ip}\cos^2\psi+e^{-iq}+\sin^3\psi]\right)
\\ \\
\left(e^{-ip}\cos^2\psi\sin\psi[e^{ip}\sin\psi-1]\right) &
\left(-\cos\psi\sin\psi[e^{-iq}+\sin\psi]\right)
\end{array} \right. \kern100pt
\\[20pt]
\left. \scriptsize
\begin{array}{c}
\left(e^{ip}\cos^4\psi\sin\psi[e^{iq}\sin\psi+1]\right) \\ \\
\left(-\cos^3\psi\sin\psi\left[e^{i(p+q)}\sin\psi+e^{ip}\sin^2\psi-e^{iq}\cos^2\psi\right]\right)
\\ \\
\left(\cos^2\psi\left[\cos^2\psi+e^{ip}\sin^2\psi\{e^{iq}+\sin\psi\}\right]\right)
\end{array} \right),
\label{rho10}
\end{multline}
\small
\begin{multline}
GF(p,\,q)E(p,\,q)=\frac{1}{\left(1+e^{iq}\varkappa\right)^2} \left(
\begin{array}{c}
\left(-e^{ip}\cos^2\psi\sin^2\psi\left[e^{iq}\sin\psi+1\right]\right)
\\ \\
\left(\cos\psi\sin^2\psi\left[e^{i(p+q)}\sin\psi+e^{ip}\sin^2\psi-e^{iq}\cos^2\psi\right]\right) \\
 \\
\left(-\sin\psi\left[\cos^2\psi+e^{ip}\sin^2\psi\{e^{iq}+\sin\psi\}\right]\right)
\end{array} \right)
\times\big(u_1,\,u_2,\,u_3\big), \kern150pt
 \\[8pt]
\big(u_1,\,u_2,\,u_3\big)\equiv
\begin{array}{cc} \bigg(
\left(\cos^4\psi-\sin\psi\left[e^{iq}\cos^2\psi+e^{i(-p+q)}\sin\psi+e^{-ip}\right]\right),
&
\left(\cos\psi\left[\cos^4\psi+\sin\psi\{e^{ip}\cos^2\psi+e^{-iq}+\sin\psi\}\right]\right),
\kern50pt
 \end{array}
   \\
\begin{array}{c}
\left(\left[e^{ip}+e^{iq}\right]\cos^4\psi\sin\psi\right)\bigg).
\end{array}
\label{rho20}
\end{multline}
\Large

Note that all columns of the matrix (\ref{rho20}) are proportional to the
third column of the matrix (\ref{rho10}).

For completeness let's give also the following formulae:
\begin{gather}
{\cal O}(p,\,q,\,r)=
\left\{e^{-ir}H+G\left(1+\frac{e^{ir}}{1+e^{ir}\sigma}F(p,\,q)\right)E(p,\,q)\right\},
\nonumber  \\[8pt]
\sigma=\frac{\left(e^{ip}+e^{iq}\right)\cos^2\psi\sin^2\psi}{1+e^{iq}\varkappa},
\qquad \varkappa=\sin^2\psi\left[e^{ip}\cos^2\psi+\sin\psi\right],
\nonumber  \\[8pt]
\chi\equiv\Big(1+e^{iq}\varkappa\Big)\Big(1+e^{ir}\sigma\Big)=1+e^{iq}\sin^3\psi+
\left[e^{i(p+q)}+e^{i(p+r)}+e^{i(q+r)}\right]\cos^2\psi\sin^2\psi,
\label{rho30}
\end{gather}
\begin{eqnarray}
H= \left(
\begin{array}{ccc}
0 & 0 &  \left(\sin\psi\right)  \\
0 & 0 &  \left(\cos\psi\sin\psi\right)   \\
0 & 0 & 0
\end{array} \right).
\label{rho40}
\end{eqnarray}

We must solve the cubic equation
\begin{gather}
\det\|{\cal O}-\rho E\|=0 \label{eig50}
\end{gather}
relative to the variable $\rho$. This equation is rewritten as
\begin{gather}
\rho^3-(\tr{\cal
O})\rho^2+\left(m_{11}+m_{22}+m_{33}\right)\rho-\det{\cal O}=0.
\label{eig60}
\end{gather}
Here $m_{ij}$ is the minor of the matrix element ${\cal O}_{ij}$.

With the help of Eqs. (\ref{rho10})-(\ref{rho40}) we find:
\begin{gather}
(\tr{\cal O})=1-\frac{\eta}{\chi}, \quad
\eta\equiv\left(1-3\cos^4\psi+\sin^6\psi\right)+\sin^3\psi\left(e^{iq}+e^{-iq}\right)=\overline{\eta}.
\label{eig70}
\end{gather}

Since the coefficients in Eq. (\ref{eig60}) are the invariants which do not
depend on the matrix ${\cal O}$ representation, we have:
\begin{gather}
\left(m_{11}+m_{22}+m_{33}\right)=\left(\rho_1\rho_2+\rho_1\rho_3+\rho_2\rho_3\right)=
\nonumber  \\[8pt]
=\left(\rho_1^{-1}+\rho_2^{-1}+\rho_3^{-1}\right)\rho_1\rho_2\rho_3=\left(\tr{\cal
O}^{\dag}\right)\det{\cal O}=\left(\overline{\tr{\cal
O}}\right)\det{\cal O}. \label{eig80}
\end{gather}
Here it was taken into account that ${\cal O}^{\dag}{\cal O}=1$. Thus it remains
to calculate the quantity $\det{\cal O}$ only.

Let $v_{\cdot i}, \,i=1,\,2,\,3$ be the vector-columns, so that
\begin{gather}
GE=\frac{1}{1+e^{iq}\varkappa}\left(v_{\cdot 1},\,v_{\cdot\, 2},
\,v_{\cdot\, 3}\right). \label{eig90}
\end{gather}
Then the matrix ${\cal O}$ is represented as follows
\begin{gather}
{\cal O}=\left(\frac{v_{\cdot 1}}{1+e^{iq}\varkappa},\,\,
\frac{v_{\cdot\, 2}}{1+e^{iq}\varkappa},\,\,w_{\cdot\, 3}\right)
-\frac{e^{ir}\sin\psi}{\cos^2\psi\left(1+e^{iq}\varkappa\right)\chi}v_{\cdot\,
3}\times\left(u_1,\,u_2,\,0\right),
\nonumber  \\[8pt]
w_{\cdot 3}=\frac{1}{\chi}v_{\cdot 3}+\tau, \quad \tau\equiv
e^{-ir}\sin\psi \left(
\begin{array}{c}
1
\\
\cos\psi
 \\
0
\end{array} \right),
\label{eig110}
\end{gather}
and $\left(u_1,\,u_2,\,u_3\right)$ is the row matrix in
(\ref{rho20}).

Let's add to the first two columns of the matrix ${\cal O}$ such columns, proportional to
$w_3$, which cancel the second summand in Eq.
(\ref{eig110}). Thus the matrix ${\cal O}'$ is obtained with the same determinant:
\begin{gather}
\Delta{\cal
O}=\frac{e^{ir}\sin\psi}{\cos^2\psi\left(1+e^{iq}\varkappa\right)}w_{\cdot
3}\times\left(u_1,\,u_2,\,0\right),
\nonumber  \\[8pt]
 {\cal O}'\equiv{\cal
O}+\Delta{\cal O}=\left(\frac{v_{\cdot 1}}{1+e^{iq}\varkappa},\,\,
\frac{v_{\cdot\, 2}}{1+e^{iq}\varkappa},\,\,w_{\cdot\, 3}\right)+
\nonumber  \\[8pt]
+\frac{\sin^2\psi}{\cos^2\psi\left(1+e^{iq}\varkappa\right)}\left(
\begin{array}{c}
1
\\
\cos\psi
 \\
0
\end{array} \right)\times\left(u_1,\,u_2,\,0\right).
\label{eig120}
\end{gather}
Now let's subtract the first row of the matrix ${\cal O}'$ multiplied by
$(\cos\psi)$ from the second row of this matrix. As a result we obtain the matrix
${\cal O}''$ the determinant of which coincides with the determinant of matrix ${\cal O}$:
\begin{multline}
{\cal O}''=  \scriptsize   \left(
\begin{array}{cc}
\frac{\cos^4\psi-e^{iq}\cos^2\psi\sin^3\psi-e^{i(-p+q)}\sin^4\psi-e^{-ip}\sin^3\psi}{\cos^2\psi\left(1+e^{iq}\varkappa\right)}
&
\frac{\sin\psi\left(\sin\psi+e^{ip}\cos^2\psi+e^{-iq}\right)}{\cos\psi\left(1+e^{iq}\varkappa\right)}
\\ \\
-\frac{\cos\psi\left(\cos^2\psi+e^{iq}\cos^2\psi\sin\psi+e^{i(-p+q)}\sin^2\psi+e^{-ip}\sin\psi\right)}{1+e^{iq}\varkappa}
 &
\frac{\cos^2\psi-\sin^2\psi-e^{ip}\cos^2\psi\sin\psi-e^{-iq}\sin\psi}{1+e^{iq}\varkappa}
\\ \\
\frac{\cos^2\psi\sin\psi\left(\sin\psi-e^{-ip}\right)}{1+e^{iq}\varkappa}
&-\frac{\cos\psi\sin\psi\left(\sin\psi+e^{-iq}\right)}{1+e^{iq}\varkappa}
 \end{array} \right. \kern100pt
\\[20pt]
\left. \scriptsize
\begin{array}{c}
\frac{\cos^4\psi\sin\psi\left(e^{i(p+q)}\sin\psi+e^{ip}\right)+e^{-ir}\chi\sin\psi}{\chi}
  \\ \\
  -\frac{\cos^3\psi\sin\psi\left(e^{i(p+q)}\sin\psi(1+\cos^2\psi)+e^{ip}-e^{iq}\cos^2\psi\right)}{\chi}
\\ \\
\frac{\cos^2\psi\left(\cos^2\psi+e^{i(p+q)}\sin^2\psi+e^{ip}\sin^3\psi\right)}{\chi}
\end{array} \right),
\label{eig130}
\end{multline}
\Large
\begin{gather}
\quad \det{\cal O}''=\det{\cal O}. \nonumber
\end{gather}

Let $m''_{ij}$ be the minor of the matrix element ${\cal
O}''_{ij}$. We need the following minors:
\begin{gather}
m''_{13}=\frac{\cos^2\psi\sin\psi\left(e^{-ip}+e^{-iq}\right)}{1+e^{iq}\varkappa},
\nonumber \\
m''_{23}=\frac{e^{-ip}\sin^3\psi-e^{-iq}\cos^2\psi\sin\psi+e^{-i(p+q)}\sin^2\psi}{\cos\psi\left(1+e^{iq}\varkappa\right)},
\nonumber \\
m''_{33}=\frac{\cos^2\psi+e^{-ip}\sin^3\psi+e^{-i(p+q)}\sin^2\psi}{\cos^2\psi\left(1+e^{iq}\varkappa\right)}.
\label{eig140}
\end{gather}

To compute the determinant let us decompose $\det{\cal O}''$ in the last
column of the matrix ${\cal O}''$. Thus we obtain:
\begin{gather}
\det{\cal O}=\frac{\overline{\chi}}{\chi}. \label{eig150}
\end{gather}
With the help of formulae (\ref{eig70}) and (\ref{eig80}) we obtain:
\begin{gather}
\left(m_{11}+m_{22}+m_{33}\right)=\frac{\overline{\chi}-\eta}{\chi}.
\label{eig160}
\end{gather}
As a result the equation (\ref{eig60}) takes the form:
\begin{gather}
\rho^3-\left(1-\frac{\eta}{\chi}\right)\rho^2+\left(\frac{\overline{\chi}}{\chi}-
\frac{\eta}{\chi}\right)\rho-\frac{\overline{\chi}}{\chi}=
(\rho-1)\left(\rho^2+\frac{\eta}{\chi}\rho+\frac{\overline{\chi}}{\chi}\right)=0.
\label{eig170}
\end{gather}
Now we find easily all eigenvalues of the matrix ${\cal
O}$:
\begin{gather}
\rho_{1,\,2}(p,\,q,\,r)=\frac{-\eta\pm
i\sqrt{4\chi\overline{\chi}-\eta^2}}{2\chi}, \quad \rho_3=1.
\label{eig180}
\end{gather}
Note that the expression $\left(4\chi\overline{\chi}-\eta^2\right)$ which is
under the sign of square root in (\ref{eig180}) is a non-negative number. Only under this condition we have
\begin{gather}
|\rho_{1,\,2,\,3}(p,\,q,\,r)|=1. \nonumber
\end{gather}
The last equalities follow also from the unitarity condition of the matrix ${\cal
O}(p,\,q,\,r)$. Otherwise it would be
\begin{gather}
|\rho_{1,\,2}(p,\,q,\,r)|\neq 1. \nonumber
\end{gather}
The last inequalities mean the violation of the matrix ${\cal O}(p,\,q,\,r)$ unitarity.

\section{The free energy}
\setcounter{equation}{0}

According to (\ref{repr4}) and (\ref{eig180})
\begin{gather}
\tr{\cal U}= 2^{MNL/2}\prod_{p,\,q,\,r}
\left(\sqrt{\rho_1}+\sqrt{\overline{\rho}_1}\right)
\left(\sqrt{\rho_2}+\sqrt{\overline{\rho}_2}\right)=2^{MNL/2}\prod_{p,\,q,\,r}\frac{\chi+\overline{\chi}-\eta}{\sqrt{\chi\overline{\chi}}}=
\nonumber \\[8pt]
=2^{MNL/2}\prod_{p,\,q,\,r}\frac{\left(1+3\cos^4\psi-\sin^6\psi\right)+2\nu(p,\,q,\,r)
\cos^2\psi\sin^2\psi}{\sqrt{\chi\overline{\chi}}},
\nonumber \\[8pt]
\nu(p,\,q,\,r)\equiv \left[\cos(p+q)+\cos(p+r)+\cos(q+r)\right].
\label{fre10}
\end{gather}

The free energy is of interest. Up to inessential summand the free energy has the form
\begin{gather}
{\cal F}=-T\ln Z=T\frac{MNL}{16\pi^3}\int_{-\pi}^{\pi}\d
p\int_{-\pi}^{\pi}\d q\int_{-\pi}^{\pi}\d r
\left(\ln\chi+\ln\overline{\chi}\right)-
\nonumber \\[8pt]
-T\frac{MNL}{8\pi^3}\int_{-\pi}^{\pi}\d p\int_{-\pi}^{\pi}\d
q\int_{-\pi}^{\pi}\d r
\ln\bigg\{\left(1+3\cos^4\psi-\sin^6\psi\right)+2\nu\cos^2\psi\sin^2\psi\bigg\}.
\label{fre20}
\end{gather}
Here it vas taken into account that at $M,\,N,\,L\rightarrow\infty$ the substitution
\begin{gather}
\sum_{p,\,q,\,r}\rightarrow\frac{MNL}{8\pi^3}\int_{-\pi}^{\pi}\d
p\int_{-\pi}^{\pi}\d q\int_{-\pi}^{\pi}\d r. \nonumber
\end{gather}
is valid.

The first summand in the right-hand side of Eq. (\ref{fre20}) is equal to zero.
This fact is the consequence of the slack inequalities
\begin{gather}
|\sigma|\leq 1, \quad |\varkappa|\leq 1, \label{fre30}
\end{gather}
so that the equalities in (\ref{fre30}) take place only on the subset
of the zeroth measure in the space of the variables $\{p,\,q,\,r\}$. Indeed,
according to (\ref{rho30}) and (\ref{fre30}) we have the following chain of equalities:
\begin{gather}
\int_{-\pi}^{\pi}\d p\int_{-\pi}^{\pi}\d q\int_{-\pi}^{\pi}\d
r\ln\chi=
\nonumber \\[8pt]
=\int_{-\pi}^{\pi}\d p\int_{-\pi}^{\pi}\d q\int_{-\pi}^{\pi}\d
r\bigg\{
\ln\big(1+e^{iq}\varkappa\big)+\ln\big(1+e^{ir}\sigma\big)\bigg\}=
\nonumber \\[8pt]
=\sum_{n=1}^{\infty}\frac{(-1)^{(n-1)}}{n}\int_{-\pi}^{\pi}\d
p\int_{-\pi}^{\pi}\d q\int_{-\pi}^{\pi}\d r
\bigg\{e^{inq}[\varkappa(\psi,\,p)]^n+e^{inr}[\sigma(\psi,\,p,\,q)]^n\bigg\}=0.
\label{fre40}
\end{gather}

Prove the slack inequalities (\ref{fre30}).

According to (\ref{EV455}) $F^2=-\sigma F$. Let's take the modulo of
the matrix element $(1,\,1)$ of the last equality:
\begin{gather}
|\sigma||F_{11}|=\left|\sum_iF_{1i}F_{i1}\right|\leq
\sqrt{\left(\sum_iF_{1i}\overline{F}_{1i}\right)\left(\sum_j\overline{F}_{j1}F_{j1}\right)}=
\nonumber \\[8pt]
=\sqrt{\left(F\,F^{\dag}\right)_{1\,1}\left(F^{\dag}F\right)_{1\,1}}.
\label{fre50}
\end{gather}
We have also from the unitarity condition (\ref{EV290}) and the evident form of
the matrix $G$ (\ref{EV270})
\begin{gather}
\left(F^{\dag}F\right)_{1\,1}=1-\left(G^{\dag}G\right)_{1\,1}=\cos^2\psi\sin^2\psi.
\label{fre60}
\end{gather}
By means of the direct calculation we find also that
\begin{gather}
\left(F\,F^{\dag}\right)_{1\,1}=\frac{1+\cos^2\psi}{\cos^2\psi}\left|F_{1\,1}\right|^2.
\label{fre70}
\end{gather}
Combining all formulae (\ref{fre50})-(\ref{fre70}) we obtain:
\begin{gather}
|\sigma|\leq\sqrt{\sin^2\psi(1+\cos^2\psi)}\leq 1 \nonumber
\end{gather}

Using the evident expression (\ref{rho30}) for $\varkappa$ we find:
\begin{gather}
|\varkappa|\leq \cos^2\psi\sin^2\psi+\sin^3\psi\leq 1 \quad
\mbox{for} \quad  0\leq\psi\leq\pi. \nonumber
\end{gather}
Thus, both inequalities (\ref{fre30}) are verified, and therefore
\begin{gather}
{\cal F}=-T\frac{MNL}{8\pi^3}\int_{-\pi}^{\pi}\d
p\int_{-\pi}^{\pi}\d q\int_{-\pi}^{\pi}\d r\times
\nonumber \\[8pt]
\times\ln\bigg\{\left(1-\sin^2\psi\right)\left[\sin^4\psi+2(\nu-1)\sin^2\psi+4\right]\bigg\}.
\label{fre80}
\end{gather}

It is seen from the expression for free energy (\ref{fre80}) that the earlier separation
of the second axis in the previous formulae for the intermediate quantities,
including the eigenvalues of the matrix ${\cal O}(p,\,q,\,r)$, was an artifact:
the free energy of the system, which presents the physical interest,
is completely symmetric relative to mutual substitutions of
the all three axes.

From the definition of the parameter $\psi$ (see the second point) it is evident that this
parameter plays a part of the temperature $T$. At $T\rightarrow\infty$ the angle
$\psi\rightarrow 0$. This follows from the fact that at
$T\rightarrow\infty$ the contribution of surfaces into the partition
function (\ref{Z60}) tends to zero. That corresponds to the high-temperature
limit in the gauge compact lattice theories. It is also seen from the comparison of the partition function
(\ref{Z60}) with the corresponding quantity of the lattice gauge theory
(with the gauge group $Z_2$) that the decrease of the temperature from the infinity
up to zero means the surgeless increase of the angular parameter $\psi$ from zero
up to $(\pi/2)$:
\begin{gather}
\frac{\d\psi(T)}{\d T}<0, \quad \psi(0)=\frac{\pi}{2}, \quad
\psi(\infty)=0. \label{fre90}
\end{gather}
The answer for the important question is need:
do the phase transitions exist at the temperature reduction
from the infinity up to zero, and what is its kind?

Since the phase transition point coincides with the peculiarity
of a free energy by the temperature, so one needs look for that values of $\psi$
at which the expression (\ref{fre80}) can have a peculiarity. Just as in the
case of the two-dimensional Ising model, the only possibility to have
a peculiarity in the free energy consists in nullification of argument of the logarithm
situated under the sign of the integral over quasi-momentum. This can occur at some
values of the temperature and quasi-momenta. Then the integral over quasi-momenta
near the peculiarity gives irregular contribution into free tnergy
and other thermodynamic quantities. In the case of the two-dimensional Ising model
there is the only point in the three-dimensional manifold of the totality
variables "two quasi-momenta+temperature" in which the logarithm argument vanish: $p=q=0$ and $T=T_c$.

In the considered three-dimensional theory the situation
differs qualitatively from the indicated one in two-dimensions: for $\psi=0 \,\, (T=\infty)$
the logarithm argument in (\ref{fre80}) is positive always;
for $p=q=r=0$ the logarithm argument in (\ref{fre80}) vanish only for $\psi=\pi/2
\,\,(T=0)$. But there is a possibility of the logarithm argument nullification in
(\ref{fre80}) at finite temperatures $0<\psi<\pi/2$ and nonzero quasi-momenta.
Indeed, at the quasi-momentum variations the parameter $\nu$
varies in the following range:
\begin{gather}
-3\leq\nu\leq 3. \label{fre130}
\end{gather}
To zero the logarithm argument in (\ref{fre80}), it is necessary realization of the
condition
\begin{gather}
x_c^2+2(\nu-1)x_c+4=0,  \quad x\equiv\sin^2\psi. \label{fre140}
\end{gather}
Only the real solutions on the segment $0\leq x_c\leq 1$ are of interest.
This is possible only under the condition
\begin{gather}
-3\leq\nu\leq-\frac32. \label{fre150}
\end{gather}
The only solution has the form
\begin{gather}
x_c(\nu)=(1-\nu)-\sqrt{(1-\nu)^2-4}\, , \label{fre160}
\end{gather}
at that $\sqrt{x_c(\nu)}$ is a monotonic increasing function on the segment
(\ref{fre150}), so that
\begin{gather}
\sqrt{x_c(-3)}=\sin\psi_c=\sqrt{4-\sqrt{12}}<1 \label{fre170}
\end{gather}
is its minimum value, and
\begin{gather}
\sqrt{x_c(-3/2)}=\sin\psi_c=1 \label{fre180}
\end{gather}
is the maximum possible value.

The point $\nu=-3$ is the evolved point in the space of quasi-momenta
since it is obtained for two values of quasi-momenta $p,\,q,\,r$ only.
Indeed, for that the quasi-momenta must satisfy at least one of the following eight
equations (all signs "plus" and "minus" in the right hand sides of Eqs.
(\ref{fre100}) are mutually independent):
\begin{gather}
p+q=\pm\pi, \quad p+r=\pm\pi,  \quad q+r=\pm\pi. \label{fre100}
\end{gather}
Since the system of linear equations
(\ref{fre100}) is non-degenerate, so for each value of the right hand side
in (\ref{fre100}) there is only one solution for quasi-momenta
$p,\,q,\,r$. Taking into account the fact that the quasi-momenta
are defined modulo $2\pi$ we can reduce all eight solutions of the systems (\ref{fre100})
to two independent solutions:
\begin{gather}
p_c^{(1)}=q_c^{(1)}=r_c^{(1)}=\frac{\pi}{2},
\nonumber \\[8pt]
p_c^{(2)}=q_c^{(2)}=r_c^{(2)}=-\frac{\pi}{2}. \label{fre110}
\end{gather}

On the contrary, for all values of $\nu$ from the half-interval
\begin{gather}
-3<\nu(p,\,q,\,r)\leq-\frac32 \label{fre190}
\end{gather}
there is a whole continuum of the quasi-momenta values for which the
logarithm argument in (\ref{fre80}) vanish.

The further study of the free energy properties and the properties of
a phase transition in the considered model will be performed in an another work.

\section{Conclusion}
\setcounter{equation}{0}

In the present work the calculation method of the partition function
of the gauge system with the gauge group $Z_2$ coupled with Majorana field  on
the three-dimensional cubic lattice is suggested.
Actually the sum over closed surfaces and
the surfaces with boundary is computed. The surfaces can be with self-itersections and
their weights are proportional to the factors $\mu^{2S}$, where
$S$ is the number of the faces of the surface.
The weights of the closed surfaces are
positive always. The total sign of the surface with
boundary depends essentially on the boundary configuration
but not on the surface form with the given boundary. This sum is computed
completely, it is represented as a threefold integral over
quasi-momenta and it is the function of one parameter (the coupling
constant or the temperature).

It should be noted that there is little examples of the integrable and
"rational" (that is having physical interpretation) statistical systems
in three dimensions, showing a phase transition relative to the temperature.
Thereupon I want pay attention to the work of A.B. Zamolodchikow \cite{7}
in which the Yang--Baxter triangle equation has been generalized to the
three dimensional case where the corresponding equation is named
as tetrahedron equation.

\appendix

\section{}

It is known \cite{3} that the rotation matrix in spinor
representation (\ref{Z70}) can be expressed in the form
\begin{gather}
{\cal U}=\exp\left(\frac14\,\omega_{{\bf x},{\bf y}}\gamma_{{\bf
x}}\gamma_{{\bf y}}\right), \quad \omega_{{\bf x},{\bf
y}}=-\omega_{{\bf y},{\bf x}}, \label{repr1}
\end{gather}
and
\begin{gather}
{\cal U}^{\dag}\gamma_{{\bf x}}{\cal U}={\cal O}_{{\bf x},{\bf
y}}\gamma_{{\bf y}}, \quad {\cal O}_{{\bf x},{\bf y}}\equiv
\left(e^{\omega}\right)_{{\bf x},{\bf y}}=\delta_{{\bf x},{\bf
y}}+\omega_{{\bf x},{\bf y}}+ \frac{1}{2!}\,\omega_{{\bf x},{\bf
z}}\omega_{{\bf z},{\bf y}}+\ldots. \label{repr2}
\end{gather}
The trace of the matrix ${\cal U}$ is expressed simply through the
eigenvalues of real orthogonal matrix ${\cal O}_{{\bf x},{\bf y}}$.
Let the set of numbers
\begin{gather}
\left(\rho_1,\,\overline{\rho}_1,\,\rho_2,\,\overline{\rho}_2,\,\ldots,\,
\rho_{3MNL/2},\,\overline{\rho}_{3MNL/2}\right)=\{\rho_k,\,\overline{\rho}_k\},
\quad k=1,\,\ldots,\,3MNL/2 \label{repr3}
\end{gather}
form the complete set of eigenvalues of the matrix ${\cal O}_{{\bf
x},{\bf y}}$. Then (see Appendix A)
\begin{gather}
\tr{\cal
U}=\prod_{k=1}^{3MNL/2}\left[2\ch\left(\frac{\ln\rho_k}{2}\right)\right]=
\prod_{k=1}^{3MNL/2}\left[2\cos\left(\frac{\phi_k}{2}\right)\right]=
\prod_{k=1}^{3MNL/2}\left(\sqrt{\rho_k}+\sqrt{\overline{\rho}_k}\right),
\nonumber \\[8pt]
\rho_k=e^{i\phi_k}. \label{repr4}
\end{gather}

Let us prove the formula (\ref{repr4}).

Let $\left\{v^{(k)}_x, \,\,\overline{v^{(k)}_x}\right\},\,\,
k=1,\ldots,\,3MNL/2$, be the complete orthonormal set of
eigenvectors of the matrix ${\cal O}_{x,\,y}$, so that the
eigenvalue $\rho_k$ $\left(\overline{\rho}_k\right)$ corresponds to
the eigenvector $v^{(k)}_x$ $\left(\overline{v^{(k)}_x}\right)$.
Further also the designation
\begin{gather}
\left\{v^{(k)}_x, \,\,\overline{v^{(k)}_x}\right\}\equiv
\left\{v^a_{{\bf{x}}}\right\}, \quad a=1,\,\ldots,\,3MNL \nonumber
\end{gather}
is used. We shall consider the introduced vectors as vector-columns
and the upper indices ${}^T$ and ${}^{\dag}$ denote the
transposition and Hermitian conjugation of vectors and matrices. By
definition
\begin{gather}
v^{(k)\,T}v^{(k')}\equiv\sum_xv^{(k)}_xv^{(k')}_x, \quad
v^{(k)\,\dag}v^{(k')}\equiv\sum_x\overline{v^{(k)}_x}v^{(k')}_x.
\label{ap1}
\end{gather}
The given definitions imply the following formulas:
\begin{gather}
v^{(k)\,T}v^{(k')}=0,  \quad v^{(k)\,\dag}v^{(k')}=\delta_{k\,k'},
\label{ap2}
\end{gather}
\begin{gather}
S_{x\,a}\equiv v^a_x \,\, \mbox{or}  \,\,
S\equiv\left(v^{(1)},\,\overline{v^{(1)}},\,v^{(2)},\,\overline{v^{(2)}},\,\ldots\right),
\quad \left(S^{\dag}S\right)_{ab}=\delta_{ab}, \label{ap3}
\end{gather}
\begin{gather}
\big(S^{\dag}{\cal
{O}}S\big)_{ab}=\diag\big(\rho_1,\,\overline{\rho}_1,\,\rho_2,\,\overline{\rho}_2,\,\ldots\big)
\equiv D_{ab}. \label{ap4}
\end{gather}
It is shown in \cite{3} that
\begin{gather}
\left(S^{\dag}\omega
S\right)_{ab}=\diag\left(\ln\rho_1,\,-\ln\rho_1,\,\ln\rho_2,\,-\ln\rho_2,\,\ldots\right)
\equiv\Delta_{ab}.
\label{ap5}
\end{gather}
Due to (\ref{ap3}) and (\ref{ap5}) we have
\begin{gather}
\sum_{x,\,y}\frac14\,\xi_x\omega_{x,\,y}\xi_y=
\sum_{x,\,y}\sum_{a,\,b}\frac14\left(\xi_xS_{x\,a}\right)\Delta_{ab}
\left(S^{\dag}_{b\,y}\xi_y\right). \label{ap6}
\end{gather}
$2^{3MNL/2}\times2^{3MNL/2}$-matrixes
\begin{gather}
c^{\dag}_k\equiv\gamma_{{\bf {x}}}v^{(k)}_{{\bf{x}}}, \quad
c_k\equiv\gamma_{{\bf {x}}}\overline{v^{(k)}_{{\bf{x}}}} \label{ap7}
\end{gather}
possess all properties of fermion creation and annihilation
operators. Indeed, in consequence of (\ref{Z10}) and (\ref{ap2})
\begin{gather}
[c_k,\,c^{\dag}_{k'}]_+=\delta_{kk'}, \quad
[c_k,\,c_{k'}]_+=[c^{\dag}_k,\,c^{\dag}_{k'}]_+=0. \label{ap8}
\end{gather}
According to the definitions (\ref{ap3}) and (\ref{ap7}) we have
\begin{gather}
\sum_x\xi_xS_{x\,a}=
\left(c^{\dag}_1,\,c_1,\,\ldots,\,c^{\dag}_{3MNL/2},\,c_{3MNL/2}\right).
 \label{ap9}
\end{gather}
With the help of Eqs. (\ref{ap5}), (\ref{ap8}) and (\ref{ap9}) the
quantity (\ref{ap6}) is rewritten as
\begin{gather}
\sum_{x,\,y}\frac14\,\xi_x\omega_{x,\,y}\xi_y=
\frac12\sum_{k=1}^{3MNL/2}\left[\ln\rho_k\left(c^{\dag}_kc_k-c_kc^{\dag}_k\right)\right]=
\sum_{k=1}^{3MNL/2}\left[\left(\ln\rho_k\right)c^{\dag}_kc_k-\frac12\ln\rho_k\right].
\label{ap10}
\end{gather}
Equality (\ref{repr4}) follows immediately from (\ref{ap10}) since
the calculation of the trace in terms of $\xi$-matrixes is
equivalent to the calculation of trace in terms of the corresponding
fermionic operators (\ref{ap7}).

\section{}

Let us consider he Majorana spinors on the three-dimensional
cubic lattice an their contribution into the partition function.

In the simplest case the action of the Dirac fermions on
the cubic lattice has the form
\begin{gather}
{\cal S}_D=\frac{i}{2}\sum_{{\bf x}}\sum_{i=1}^3\overline{\psi}_{\bf
x}\gamma^i\left(U_{{\bf x},\,{\bf e}_i}\psi_{{\bf x}+{\bf
e}_i}-U^{\dag}_{{\bf x}-{\bf e}_i,\,{\bf e}_i}\psi_{{\bf x}-{\bf
e}_i}\right),
\nonumber \\[8pt]
\gamma^1=\sigma_x, \quad \gamma^2=\sigma_y,  \quad
\gamma^3=\sigma_z,   \quad   U^{\dag}_{{\bf x},\,{\bf e}_i}U_{{\bf
x},\,{\bf e}_i}=1.
\label{apb10}
\end{gather}
Here $\sigma_i$ are the Pauli matrixes and $U_{{\bf x},\,{\bf e}_i}$ is the gauge field.
The Fermi-fields $\psi_{{\bf x}}$ and
$\overline{\psi}_{{\bf x}}$ are the elements of the Grassman algebra
and all their elements are considered as a mutually independent variables.
The Fermi contribution to the partition function is defined as the integral
\begin{gather}
Z_D\{U\}=\prod_{\bf x}\int\d\overline{\psi}_{{\bf x}}\d\psi_{{\bf
x}}\exp{\cal S}_D,
\label{apb20}
\end{gather}
where $\d\overline{\psi}_{{\bf x}}$ and $\d\psi_{{\bf x}}$ denote the products
of the differentials of the both components of the corresponding spinors.

The Majorana spinors are determined by the following
system of identifications:
\begin{gather}
\overline{\psi}_{{\bf x}}=-\psi^T_{\bf x}\gamma^2.
\label{apb30}
\end{gather}
The "electrical current" of the Majorana spinors is equal to zero identically:
\begin{gather}
J^i_{\bf x}=\overline{\psi}_{\bf x}\gamma^i\psi_{\bf x}=
\nonumber \\[8pt]
=(\psi_{{\bf x}1},\,\psi_{{\bf x}2}) \left(
\begin{array}{cc}
0 &  i  \\
-i & 0
\end{array} \right)
\left\{\left(
\begin{array}{cc}
0 &  1  \\
1 & 0
\end{array} \right),\,\,
\left(
\begin{array}{cc}
0 &  -i  \\
i & 0
\end{array} \right),\,\,
\left(
\begin{array}{cc}
1 &  0  \\
0 & -1
\end{array} \right)\right\}\left(
\begin{array}{c}
\psi_{{\bf x}1} \\
\psi_{{\bf x}2}
\end{array} \right)=
\nonumber \\[8pt]
=\left\{i(\psi_{{\bf x}1}^2-\psi_{{\bf x}2}^2),\,-(\psi_{{\bf
x}1}^2+\psi_{{\bf x}2}^2),\,-i(\psi_{{\bf x}1}\psi_{{\bf
x}2}+\psi_{{\bf x}2}\psi_{{\bf
x}1})\right\}\equiv\left\{0,\,0,\,0\right\}.
\label{apb40}
\end{gather}
The last identity in (\ref{apb40}) follows from the fact
that all components of the field $\psi$ are odd elements of the Grassman algebra.
The majorana action is obtained from the Dirac action (\ref{apb10})
by means of the substitutions $\overline{\psi}_{\bf x}\rightarrow-\psi^T_{\bf
x}\gamma^2$ and division by 2:
\begin{gather}
{\cal S}_M=-\frac{i}{4}\sum_{{\bf x}}\sum_{i=1}^3\psi^T_{\bf
x}\gamma^2\gamma^i\left(U_{{\bf x},\,{\bf e}_i}\psi_{{\bf x}+{\bf
e}_i}-U^{\dag}_{{\bf x}-{\bf e}_i,\,{\bf e}_i}\psi_{{\bf x}-{\bf
e}_i}\right).
\label{apb50}
\end{gather}
The contribution of Majorana spinors to the partition function
is determined by the Grassman integral
\begin{gather}
Z_M\{U\}=\prod_{\bf x}\int\d\psi_{{\bf x}2}\d\psi_{{\bf
x}1}\exp{\cal S}_M.
\label{apb60}
\end{gather}
Since
\begin{gather}
\int\d\psi_{{\bf x}2}\d\psi_{{\bf x}1}\left(\psi_{{\bf
x}1}\psi_{{\bf x}2}\right)^n=\left\{
\begin{array}{cl}
1, &  \mbox{for} \ \ n=1 \\[5pt]
0,  & \mbox{for} \ \  n\neq 1
\end{array} \right.,
\nonumber
\end{gather}
so
\begin{gather}
\int\d\psi_{\bf x}=0,
\nonumber \\[8pt]
-i\int\d\psi_{\bf x}\cdot\left(\psi_{\bf x}\psi_{\bf
x}^T\right)=-i\int\d\psi_{{\bf x}2}\d\psi_{{\bf x}1}\left(
\begin{array}{cc}
0 &  \psi_{{\bf
x}1}\psi_{{\bf x}2}  \\
\psi_{{\bf x}2}\psi_{{\bf x}1} & 0
\end{array} \right)=\gamma^2,
\nonumber \\[8pt]
\int\d\psi_{\bf x}\cdot\left(\psi_{\bf x}\psi_{\bf
x}^T\right)\otimes\left(\psi_{\bf x}\psi_{\bf x}^T\right)=0,
\label{apb70}
\end{gather}
and so on.

Draw on the lattice the system of closed broken contours without
intersections and self-intersections. The elementary link of each
contour is an edge $\ml_{{\bf x},\,i}$ (see Section 2.2) and each
vertex belong to one and only one contour. Each contour is oriented,
that is the direction (arrow) is assigned to each edge of the
contour, so that the continuous movement along the arrows reduce to
the whole round of the contour. We shall call the edge as positive
(negative) oriented if its arrow is directed in the line of the
positive (negative) direction of the corresponding lattice axis.
Below the particular case of the elementary closed contour based on
only one edge with both orientations at once is described.

The nonzero contribution in the integral (\ref{apb60}) give only
those summands in the exponent expansion in the quantities
\begin{gather}
\left(-\frac{i}{4}\psi^T_{\bf x}\gamma^2\gamma^iU_{{\bf x},\,{\bf
e}_i}\psi_{{\bf x}+{\bf e}_i}\right)
\label{apb80}
\end{gather}
and
\begin{gather}
\left(\frac{i}{4}\psi^T_{\bf x}\gamma^2\gamma^iU^{\dag}_{{\bf
x}-{\bf e}_i,\,{\bf e}_i}\psi_{{\bf x}-{\bf e}_i}\right),
\label{apb90}
\end{gather}
in which these quantities are in the first power. The factor
(\ref{apb80}) corresponds to each positively oriented edge and the
factor (\ref{apb90}) corresponds to each negatively oriented edge
\footnote{This statement is wrong in the case of the gauge group
$Z_2$, but in the Appendix B the others gauge group are meant.}. The
both factors (\ref{apb80}) and (\ref{apb90}) correspond to the
elementary closed contour based on the edge $\ml_{{\bf x},\,i}$.
Thereby, under the sign of the integral in each vertex ${\bf x}$
there is the element of the Grassman algebra $\left(\psi_{\bf
x}\psi_{\bf x}^T\right)$ in the first power giving a nonzero factor
according to (\ref{apb70}). Note that the linkage of this rule with
the contours orientation in the Majorana case is due to the
interaction between fermions and gauge field. Indeed, the
calculation of the integral over the gauge field in the frame of the
high temperature expansion leads to the nullification of all
contributions from the non-oriented closed contours (in the case of
the higher gauge symmetry than $Z_2$) which do not vanish as a
result of the fermion integration (\ref{apb60}).

Here the specific question is interesting for us. Therefore further
we put $U_{{\bf x},\,{\bf e}_i}=1$.

From the aforesaid and with the help of Eqs. (\ref{apb50}),
(\ref{apb70}), (\ref{apb80}) and (\ref{apb90}) the following rules
follow for the calculation of the integral (\ref{apb60}):

1) Let's draw on the lattice the system of the outlined above closed
broken oriented lines and, rounding each contour successively along
the arrows, relate successively to each edge $\ml_{{\bf x},\,i}$ the
factor $\left(1/4\right)\gamma^i$ in the case of the positive
orientation of the edge and the factor $\left(-1/4\right)\gamma^i$
in the case of the negative orientation of the edge.

2) After ending the rounding process of each contour let's calculate
the trace of the ordered product of the $\gamma$-matrixes
corresponding to the contour according to the rule 1, and add the
factor $(-1)$, common for the whole contour. The obtained number is
called as the factor of the contour.

3) Let's multiply the factors of all contours. The obtained number
is called as the factor of the system of contours.

4) For finding the integral (\ref{apb60}) it is necessary to
summarize the factors of all possible systems of contour.

Now let's show the change of the sign of the contour system factors
on the simplest examples. Since only the sign is interesting for us,
the others positive factors are ignored.

Let's consider the factor of the elementary closed contour based on
the edge $\ml_{{\bf x},\,i}$. This contour is represented in the
Fig. 4.

\psfrag{vx}{\rotatebox{0}{\kern0pt\lower0pt\hbox{{$\ve{x}$}}}}
\psfrag{f401}{\rotatebox{0}{\kern0pt\lower0pt\hbox{{$\ve{x}+\ve{e}_i$}}}}
\begin{center}
\includegraphics[scale=0.7]{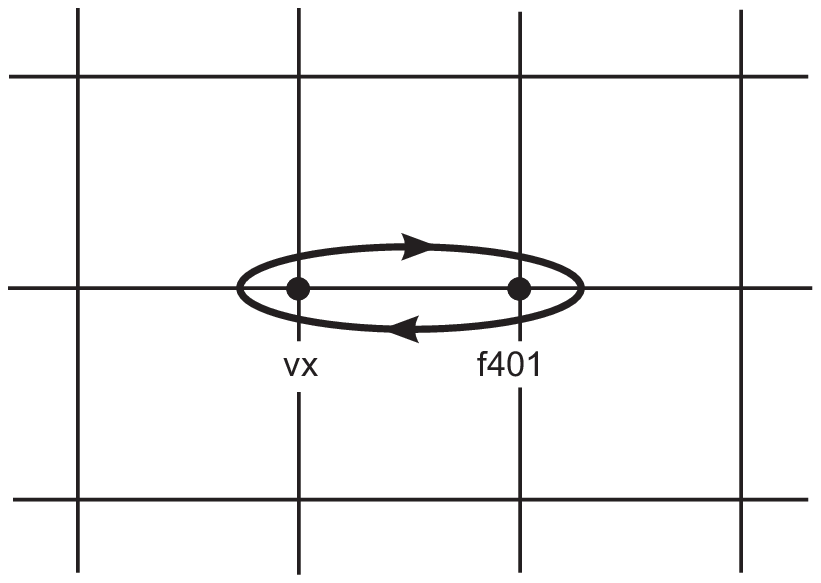}
\end{center}
\vskip8pt

\begin{center}
FIG. 4
\end{center}
\vskip8pt


According to the given rules the factor of this contour is
\begin{gather}
\Phi_{{\bf x},\,i}=(-1)\tr\gamma^i(-\gamma^i)=1.
\label{apb100}
\end{gather}
One of the possible and at the same time simplest system of contours
is represented in the Fig. 5. In Fig. 5 one of the mutually parallel
planes containing the base vector ${\bf e}_i$ is represented. Thus,
all these planes contain the identical configurations of elementary
closed contours and each vertex of the lattice belong to one and
only one contour. It is evident that due to (\ref{apb100}) the
factor corresponding to this system of contours is equal to the
product of units and thus it is equal to the unity.

\psfrag{f501}{\rotatebox{0}{\kern0pt\lower0pt\hbox{{$\ve{x}+\ve{e}_i$}}}}
\begin{center}
\includegraphics[scale=0.7]{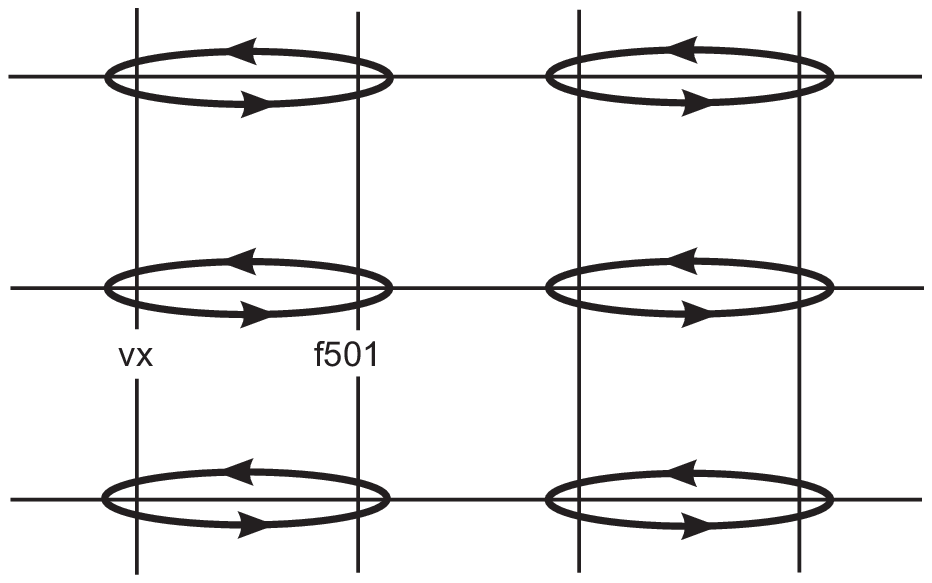}
\end{center}
\vskip8pt

\begin{center}
FIG. 5
\end{center}
\vskip8pt


It is supposed that in the subsequent examples of the closed contour
systems almost all closed contours are elementary, so that the
change of the number of the elementary contours does not affect on
the total sign of the contour system. Therefore only that closed
contours will be considered and represented in the figures which can
effect on the total sign of the contour system factor.

Let's consider the factor corresponding to the closed contour in the
plane $(1,\,2)$ which is represented in the Fig. 6.

\psfrag{f0601}{\rotatebox{0}{\kern0pt\lower0pt\hbox{{$\ve{x}$}}}}
\psfrag{f0602}{\rotatebox{0}{\kern0pt\lower0pt\hbox{{$\ve{x}+\ve{e}_1$}}}}
\psfrag{f0603}{\rotatebox{0}{\kern0pt\lower0pt\hbox{{$\ve{x}+\ve{e}_2$}}}}
\psfrag{f0604}{\rotatebox{0}{\kern0pt\lower0pt\hbox{{$\ve{x}+\ve{e}_1+\ve{e}_2$}}}}
\begin{center}
\includegraphics[scale=0.7]{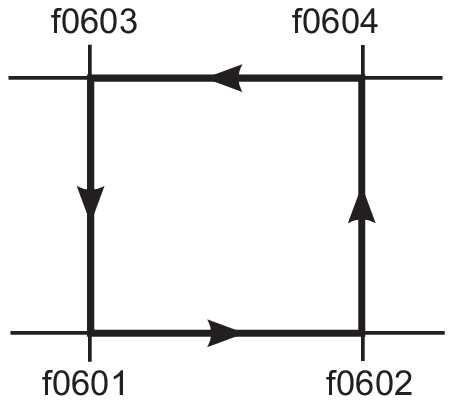}
\end{center}
\vskip8pt

\begin{center}
FIG. 6
\end{center}
\vskip8pt


The factor corresponding to the contour in the Fig. 6 is
\begin{gather}
\Phi_{{\bf x},\,{\bf e}_1,\,{\bf
e}_2}=(-1)\tr\gamma^1\gamma^2(-\gamma^1)(-\gamma^2)=1.
\label{apb110}
\end{gather}
Therefore the total sign of the contour system factor corresponding
to the Fig. 6 is also equal to unity.

Now let's consider the factor corresponding to the closed contour in
the plane $(1,\,2)$ and represented in Fig. 7.

\psfrag{f0701}{\rotatebox{0}{\kern0pt\lower0pt\hbox{{$\ve{x}$}}}}
\psfrag{f0702}{\rotatebox{0}{\kern0pt\lower0pt\hbox{{$\ve{x}+\ve{e}_1$}}}}
\psfrag{f0703}{\rotatebox{0}{\kern0pt\lower0pt\hbox{{$\ve{x}+2\ve{e}_1$}}}}
\psfrag{f0704}{\rotatebox{0}{\kern0pt\lower0pt\hbox{{$\ve{x}+2\ve{e}_1+\ve{e}_2$}}}}
\psfrag{f0705}{\rotatebox{0}{\kern0pt\lower0pt\hbox{{$\ve{x}+2\ve{e}_1+2\ve{e}_2$}}}}
\psfrag{u1}{\rotatebox{0}{\kern0pt\lower0pt\hbox{{$\ve{x}+\ve{e}_1+\ve{e}_2$}}}}
\psfrag{f0707}{\rotatebox{0}{\kern0pt\lower0pt\hbox{{$\ve{x}+\ve{e}_2$}}}}
\psfrag{f0708}{\rotatebox{0}{\kern0pt\lower0pt\hbox{{$\ve{x}+\ve{e}_2$}}}}
\begin{center}
\includegraphics[scale=0.7]{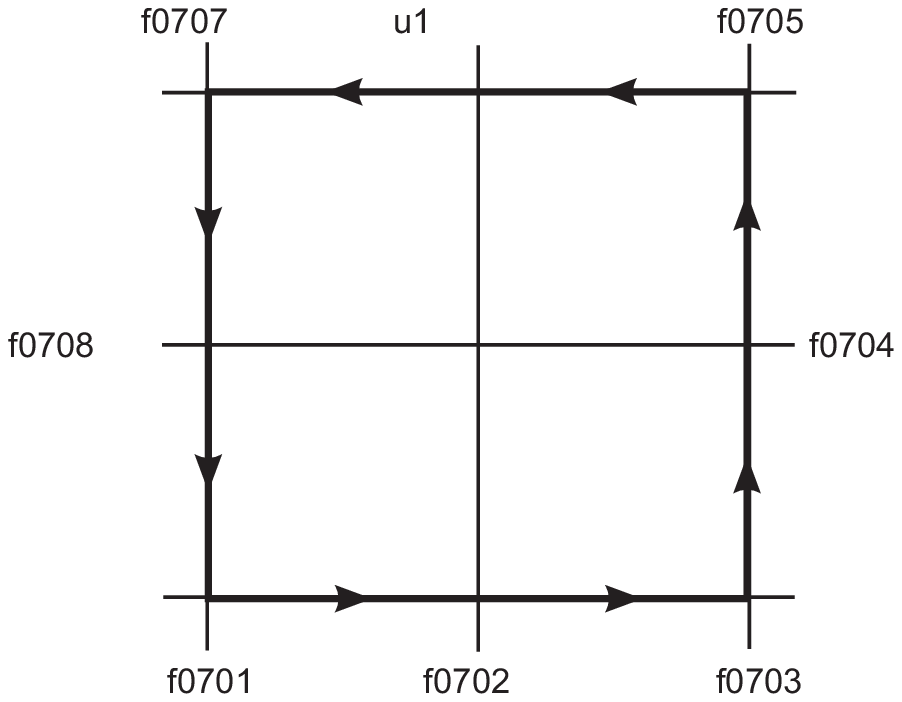}
\end{center}
\vskip8pt

\begin{center}
FIG. 7
\end{center}
\vskip8pt


\begin{gather}
\Phi_{{\bf x},\,2{\bf e}_1,\,2{\bf
e}_2}=(-1)\tr\gamma^1\gamma^1\gamma^2\gamma^2(-\gamma^1)(-\gamma^1)(-\gamma^2)(-\gamma^2)=-1.
\label{apb120}
\end{gather}
From here it is seen that the total sign of the contour system
factor corresponding to the Fig. 7 is negative.

Thus we see that the contour system factor, the sum of which defines
the integral (\ref{apb60}), can have either positive or negative
sign. It is important that the change of the contour configurations
leads, generally speaking, to the change of the factor sign.

In conclusion note that the fermion {\it Dirac} contribution into
the partition function (\ref{apb20}) is expressed also as a sum of
the contour system factor. But in the Dirac case the sign of the
factor does not depend on the contour configurations. It is easy to
see that in the long-wave continuous limit in a weak gauge field. In
this limit the lattice action (\ref{apb10}) transforms into the
usual Dirac action
\begin{gather}
{\cal
S}_D=\int\d^3x\overline{\psi}\left(i\gamma^i\partial_i-e\gamma^iA_i\right)\psi.
\label{apb130}
\end{gather}
The corresponding contribution into the partition function can be
writthen in the form
\begin{gather}
\det\left(i\gamma^i\partial_i-e\gamma^iA_i\right)=\mbox{Const}
\cdot\exp\left\{\tr\ln\left[1-e(i\gamma^i\partial_i)^{-1}\gamma^jA_j\right]\right\}=
\nonumber \\[8pt]
=\mbox{Const}\cdot\exp\left\{-e^2\int\d^3x\d^3y\tr
(i\gamma^k\partial_k)^{-1}_{x,\,y}\gamma^iA_i(y)(i\gamma^k\partial_k)^{-1}_{y,\,x}\gamma^jA_j(x)-\ldots\right\}.
\label{apb140}
\end{gather}
In the last expansion in the exponent under integrals the enough
smooth function $(i\gamma^k\partial_k)^{-1}_{x,\,y}$ is present.
Since the space integrations mean the contour variations, so it is
seen that the contour variations is not conjugated with the sign
variation of the corresponding factor. The variations of the contour
factor signs of the lattice Majorana fermions mean that in the
continuous long-wave limit the contributions into the partition
function from these contours are cancelled mutually. From here the
impossibility of Majorana fermions-Abelian gauge field interaction
in the case of continuous theory is seen. This state also follows
directly from (\ref{apb40}).

\end{document}